\documentclass[a4paper, 11pt]{article}
\usepackage{jheppub}

\usepackage[
    colorlinks = true, linktocpage = true,
    linkcolor = blue, citecolor = blue, 
    urlcolor = blue]{hyperref}

\usepackage[utf8]{inputenc}
\usepackage[T1]{fontenc}
\usepackage{amsmath}
\usepackage{graphicx}
\usepackage{slashed}
\usepackage{booktabs}

\newcommand{\IPB}{I^\mathrm{PB}}  % pentabox integral
\newcommand{\IHB}{I^\mathrm{HB}}  % hexabox integral
\newcommand{\nr}{R}  % nested root
\renewcommand{\Re}{\mathrm{Re}}
\renewcommand{\Im}{\mathrm{Im}}

\title{Two Feynman integral families for vector boson fusion at NNLO QCD}

\author{Linus Götzfried}
\author{and Andreas von Manteuffel}
\affiliation{Institute for Theoretical Physics, 
    University of Regensburg, 93053 Regensburg, Germany}

\emailAdd{linus.goetzfried@ur.de}
\emailAdd{manteuffel@ur.de}

\abstract{We consider a planar and a non-planar two-loop Feynman integral family arising in non-factorisable quantum chromodynamics corrections to Higgs-boson production via vector-boson fusion.
We construct basis integrals which fulfil canonical differential equations with respect to seven variables for the external kinematics and an internal mass.
One of the integral families involves nested square roots.
We present strategies for deriving canonical master integrals, $\mathrm{d}\log$ forms and analytic continuation for such cases.
Including also families with crossed massless legs, we express the $\varepsilon$-expanded integrals in terms of an algebraically independent function basis.
For their evaluation, we employ numerical integration of the differential equations.
}

\begin{document}

\maketitle

%%%%%%%%%%%%%%%%%%%%%%%%%%%%%%%%%%%%%%%%%%%%%%%%%%

\section{Introduction}
Vector boson fusion (VBF) is the second most dominant mode of Higgs-boson production at the Large Hadron Collider (LHC). 
A distinctive experimental signature of two associated high-energetic forward-directed jets make it interesting for precision studies of the Standard Model and searches for deviations. 
Using strict kinematical cuts, it can be cleanly separated from other processes with the same initial and final states, e.g.\ Higgs-Strahlung (VH) with hadronic decay. 
Since VBF directly probes the electroweak $HVV$-vertex, it is in particular sensitive to anomalous couplings of the Higgs boson to electroweak vector bosons.

\begin{figure}[t]
\centering
\begin{tabular}{ll}
\includegraphics[scale = 0.75]{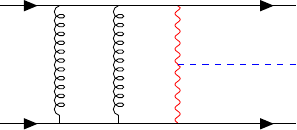} & 
\includegraphics[scale = 0.75]{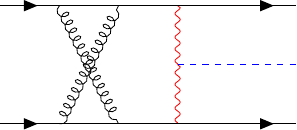} \\[5pt]
\includegraphics[scale = 0.75]{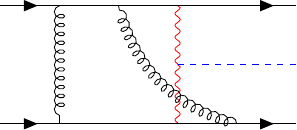} & 
\includegraphics[scale = 0.75]{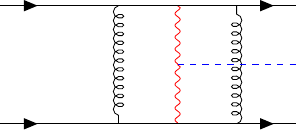}
\end{tabular}
\caption{Representative two-loop Feynman diagrams for irreducible corrections to vector boson fusion at NNLO QCD. 
Solid lines depict quarks, curly lines gluons, wiggly lines (red) weak vector bosons and the dashed line (blue) the Higgs boson.
This work considers the two topologies shown in the first row.}
\label{fig:VBF2Loop}
\end{figure}

Future LHC measurements will both improve experimental precision and explore the regime of relaxed VBF cuts~\cite{Berger:2019wnu}, where clean separation from VH production can no longer be guaranteed.
While for conventional VBF cuts the fiducial cross section is dominated by well-understood factorisable contributions without colour exchange between the two incoming partons~\cite{Figy:2003nv, Figy:2004pt, Berger:2004pca, Ciccolini:2007ec, Figy:2010ct, Bolzoni:2010xr, Bolzoni:2011cu, Cacciari:2015jma, Dreyer:2016oyx, Cruz-Martinez:2018rod}, this is not as clear once the cuts are relaxed.
This motivates a calculation of the non-factorisable quantum chromodynamics (QCD) corrections, which arise first at next-to-next-to-leading order (NNLO).
They are significantly more complex than the factorisable contributions, since they involve the full five-point kinematics of the VBF process. 
Representative Feynman diagrams are drawn in Figure~\ref{fig:VBF2Loop}, depicting examples for the four different integral families arising in the computation.

Due to the lack of complete two-loop amplitudes, an exact NNLO prediction is not yet available in the literature.
Existing NNLO QCD predictions for non-factorisable corrections have been performed within the eikonal approximation~\cite{Liu:2019tuy, Long:2023mvc, Asteriadis:2023nyl}, which is sufficient for conventional VBF cuts but not necessarily for more general regions of phase space~\cite{Asteriadis:2023nyl, Dreyer:2020urf}.

A major obstacle for the exact two-loop computation are the complicated five-point Feynman integrals appearing in the non-factorisable topologies shown in Figure~\ref{fig:VBF2Loop}.
In this work, we consider the planar penta-box and a the non-planar hexa-box shown in the first row of Figure~\ref{fig:VBF2Loop}. 
We compute them using the method of canonical differential equations.

The presence of an internal and an external mass scale make this problem a seven-variable one, similar to several recent two-loop computations (e.g.~\cite{FebresCordero:2023pww, Abreu:2024yit, Becchetti:2025qlu}).
We find that, although the basis change to the canonical basis is algebraic for both involved topologies, one subsector of the penta-box requires the introduction of nested square roots.
Nested square roots have been encountered in the study of a similar topology for $t\overline{t}H$ production~\cite{FebresCordero:2023pww}, where not all differential forms were expressed as $\mathrm{d}\log$ forms.\footnote{Note that the subtopology involving nested square roots appears also in electroweak corrections to $gg\rightarrow ZH$~\cite{Li:2026emp}.}
Whether this is possible or not is not straight-forward to observe, since the presence of nested square roots complicates subsequent algebra and the derivation of differential equations significantly.
Nevertheless, for the topology considered in this work, we find that the differential equations can be expressed entirely in $\mathrm{d}\log$ form. 
We describe in detail the strategy to determine the corresponding letters, which we believe to be useful for similar problems.
For the hexa-box topology, difficulties arise mostly from its non-planarity, leading to more complex expressions.
We consider the master integrals for arbitrary permutations of the massless legs and express their $\varepsilon$-expansion in terms of a common basis of algebraically independent functions.
For their evaluation, we employ numerical integration of the differential equations.

Our paper is structured as follows: 
We begin by introducing relevant kinematics and integral families in Section~\ref{sec:KinematicsFamilies}. 
In Section~\ref{sec:EpsilonBasis}, we describe our canonical basis and the strategies used to find it, and in Section~\ref{sec:Alphabet} we elaborate on the symbol alphabet of the integral families and its derivation. 
Section~\ref{sec:DeqsPermutations} describes how we find the differential equations and how we transform them for permutations of the massless external legs. 
In Section~\ref{sec:FunctionBasis}, we describe the expression in terms of a transcendental function basis and present numerical results.
We conclude in Section~\ref{sec:Conclusions}.

\section{Kinematics and integral families}
\label{sec:KinematicsFamilies}
\begin{figure}[t]
\centering
\includegraphics[scale = 0.75]{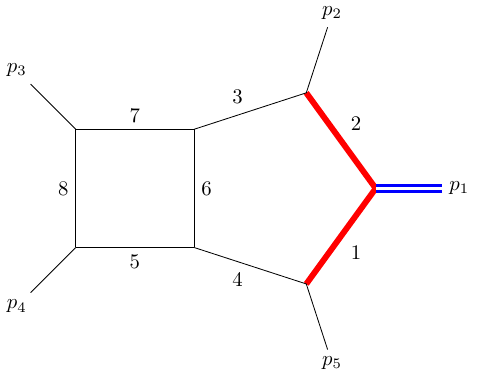}
\hspace{0.5cm}
\raisebox{0.42cm}{
\includegraphics[scale = 0.75]{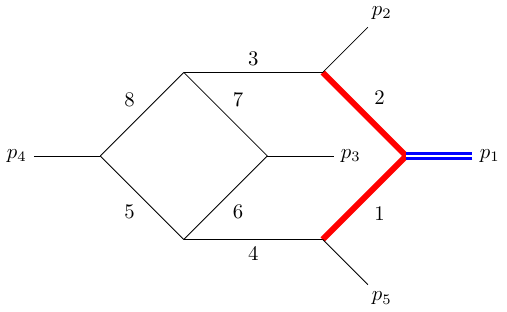}}
\caption{Two-loop Feynman integral families considered in this work. 
Left: penta-box ``PB'', right: hexa-box ``HB''. 
Thick lines (red) indicate internal particles with mass $m$, double lines (blue) indicate external particles with momentum square $s_1$, and all other lines are massless. 
Moreover, we enumerate the internal lines and depict the external momenta, all of which are chosen incoming.}
\label{fig:Families}
\end{figure}

We consider two five-point Feynman integral families, a penta-box ``PB'' and a hexa-box ``HB'', as shown in Figure~\ref{fig:Families}.
Explicit definitions of the families are
\begin{gather}
    \IPB_{\nu_1\dots\nu_{11}} \equiv e^{2\varepsilon\gamma_E} \mu_0^{4\varepsilon}
    \int \frac{\mathrm{d}^d k_1}{i\pi^{d/2}} \frac{\mathrm{d}^d k_2}{i\pi^{d/2}} 
    \prod_{a = 1}^{11} \frac{1}{(D_a^\mathrm{PB})^{\nu_a}} \, , \nonumber\\
    \begin{aligned}[b]
    D_1^\mathrm{PB} & \equiv k_1^2 - m^2 \, , & 
    D_2^\mathrm{PB} & \equiv (k_1 - p_2 - p_3 - p_4 - p_5)^2 - m^2 \, , \\
    D_3^\mathrm{PB} & \equiv (k_1 - p_3 - p_4 - p_5)^2 \, , & 
    D_4^\mathrm{PB} & \equiv (k_1 - p_5)^2 \, , \\
    D_5^\mathrm{PB} & \equiv k_2^2 \, , & D_6^\mathrm{PB} & 
    \equiv (k_2 - k_1 + p_5)^2 \, , \\
    D_7^\mathrm{PB} & \equiv (k_2 - p_3 - p_4)^2 \, , & 
    D_8^\mathrm{PB} & \equiv (k_2 - p_4)^2 \, , \\
    D_9^\mathrm{PB} & \equiv (k_1 - p_4 - p_5)^2 \, , & 
    D_{10}^\mathrm{PB} & \equiv (k_2 - p_3 - p_4 - p_5)^2 \, , \\
    D_{11}^\mathrm{PB} & \equiv (k_2 - p_2 - p_3 - p_4 - p_5)^2 \, ,
    \end{aligned}
\label{eq:Families:PB}
\end{gather}
with $\nu_9,\nu_{10},\nu_{11}\leq 0$, and
\begin{gather}
    \IHB_{\nu_1\dots\nu_{11}} \equiv e^{2\varepsilon\gamma_E} \mu_0^{4\varepsilon}
    \int \frac{\mathrm{d}^d k_1}{i\pi^{d/2}} \frac{\mathrm{d}^d k_2}{i\pi^{d/2}} 
    \prod_{a = 1}^{11} \frac{1}{(D_a^\mathrm{HB})^{\nu_a}} \, , \nonumber\\
    \begin{aligned}[b]
    D_1^\mathrm{HB} & \equiv k_1^2 - m^2 \, , & 
    D_2^\mathrm{HB} & \equiv (k_1 - p_2 - p_3 - p_4 - p_5)^2 - m^2 \, , \\
    D_3^\mathrm{HB} & \equiv (k_1 - p_3 - p_4 - p_5)^2 \, , & 
    D_4^\mathrm{HB} & \equiv (k_1 - p_5)^2 \, , \\
    D_5^\mathrm{HB} & \equiv k_2^2 \, , & 
    D_6^\mathrm{HB} & \equiv (k_2 - k_1 + p_5)^2 \, , \\
    D_7^\mathrm{HB} & \equiv (k_2 - k_1 + p_4 + p_5)^2 \, , & 
    D_8^\mathrm{HB} & \equiv (k_2 - p_3)^2 \, , \\
    D_9^\mathrm{HB} & \equiv (k_1 - p_4 - p_5)^2 \, , & 
    D_{10}^\mathrm{HB} & \equiv (k_2 - p_2 - p_3)^2 \, , \\
    D_{11}^\mathrm{HB} & \equiv (k_2 - p_2 - p_3 - p_4 - p_5)^2 \, , 
    \end{aligned}
\label{eq:Families:HB}
\end{gather}
with $\nu_9,\nu_{10},\nu_{11}\leq 0$. 
In both cases, the inverse propagators $9$, $10$ and $11$ are irreducible scalar products and may only arise in the numerator. 
The space-time dimension is $d = 4 - 2 \, \varepsilon$, $\gamma_E$ is the Euler-Mascheroni constant and $\mu_0$ is an arbitrary scale introduced such that the integrals have integer mass dimension also for $d\neq 4$.

All external momenta $p_i$, $i=1,\ldots,5$ are chosen incoming, with
\begin{align}
    p_1^2 \equiv s_1 \, , \qquad 
    p_i^2 = 0 \, \, \textnormal{for} \, i\neq 1 \, , \qquad 
    \sum_{i = 1}^5 p_i = 0 \, .
\end{align}
The internal mass is denoted by $m$ and $s_{ij} \equiv (p_i + p_j)^2$ are Mandelstam invariants. 
Then, scattering kinematics can be described in terms of a vector of seven independent Lorentz invariants,
\begin{align}
\label{eq:Kinematics:Vector}
    \vec{x} \equiv (m^2,s_1,s_{12},s_{23},s_{34},s_{45},s_{15})^T \, , 
\end{align}
and the sign of the parity-odd invariant $\mathrm{tr}_5 \equiv \mathrm{tr}(\slashed{p}_2\slashed{p}_3\slashed{p}_4\slashed{p}_5\gamma_5)$.
We define (generalised) Gram determinants
\begin{align}
    G(\{q_1,\dots,q_n\},\{q_1^\prime,\dots,q_n^\prime\}) & \equiv \det((q_i\cdot q_j^\prime)_{1\leq i,j\leq n}) \, , \nonumber\\
    G(\{q_1,\dots,q_n\}) & \equiv G(\{q_1,\dots,q_n\},\{q_1,\dots,q_n\}) \, .
\end{align}
Then, $\mathrm{tr}_5^2 = 16 \, \Delta_5$, where
\begin{align}
    \Delta_5 & \equiv G(\{p_2,p_3,p_4,p_5\}) \nonumber\\
    & = \frac{1}{16} \, (s_{12}^2 s_{15}^2 - 2 \, s_{12}^2 s_{15} s_{23} + s_{12}^2 s_{23}^2 - 2 \, s_1 s_{12} s_{15} s_{34} + 2 \, s_1 s_{12} s_{23} s_{34} \nonumber\\
    & \hspace{11pt} + 2 \, s_{12} s_{15} s_{23} s_{34} - 2 \, s_{12} s_{23}^2 s_{34} + s_1^2 s_{34}^2 - 2 \, s_1 s_{23} s_{34}^2 + s_{23}^2 s_{34}^2 - 2 \, s_{12} s_{15}^2 s_{45} \nonumber\\
    & \hspace{11pt} + 2 \, s_{12} s_{15} s_{23} s_{45} + 2 \, s_1 s_{15} s_{34} s_{45} + 2 \, s_{12} s_{15} s_{34} s_{45} - 4 \, s_1 s_{23} s_{34} s_{45} \nonumber\\
    & \hspace{11pt} + 2 \, s_{12} s_{23} s_{34} s_{45} + 2 \, s_{15} s_{23} s_{34} s_{45} - 2 \, s_1 s_{34}^2 s_{45} - 2 \, s_{23} s_{34}^2 s_{45} + s_{15}^2 s_{45}^2 \nonumber\\
    & \hspace{11pt} - 2 \, s_{15} s_{34} s_{45}^2 + s_{34}^2 s_{45}^2)
    \label{eq:Delta5}
\end{align}
is the five-point Gram determinant.
Since the scalar integrals considered in this work are parity even, they do not depend on $\mathrm{tr}_5$ itself and we will mostly ignore it in the following.
 
Equations~(\ref{eq:Families:PB}~-~\ref{eq:Families:HB}) define the PB and HB families in the identity permutation of the massless legs. 
In this work, we compute the integral families in all $24$ permutations of the massless legs, which allows us to eliminate redundancies between different permutations in our special function representation, cf.\ Section~\ref{sec:FunctionBasis}. 
Then, it is sufficient for the application to VBF to evaluate the integrals in one specific $2\rightarrow 3$-scattering channel (see e.g.\ ref.~\cite{Chicherin:2020oor}). 
We choose the $s_{45}$-channel with particles $4$ and $5$ incoming, which was also used in the ancillary files to ref.~\cite{Asteriadis:2026olo}.
Except for a subset of degenerate momentum configurations which are not of interest here, it is defined by the inequalities (see ref.~\cite{Chicherin:2021dyp})
\begin{gather}
    s_1 > 0 \, , \quad s_{12} > s_1 \, , \quad s_{45} > s_1 \, , \quad 
    - s_{12} - s_{23} + s_{45} > 0 \, , \quad s_{23} > 0 \, , \nonumber\\
    s_{15} - s_{23} - s_{34} < 0 \, , \quad s_{34} < 0 \, , \quad 
    s_1 - s_{12} - s_{15} + s_{34} < 0 \, , \nonumber\\
    s_{12} - s_{34} - s_{45} < 0 \, , \quad s_1 - s_{15} + s_{23} - s_{45} < 0 \, , \quad 
    s_{15} < 0 \, , \quad \Delta_5<0 \, .
\label{eq:s45Region}
\end{gather}
Additionally, we require $\Re(m^2)>0$ and $\Im(m^2)\leq 0$, where $\Im(m^2)\neq 0$ in the complex-mass scheme~\cite{Denner:1999gp}.
In the following, in Sections~\ref{sec:EpsilonBasis},~\ref{sec:Alphabet}, we first focus on the integral families in their identity permutations and consider the permutation closure later in Sections~\ref{sec:DeqsPermutations},~\ref{sec:FunctionBasis}.

\section{Determination of a basis with \texorpdfstring{\boldmath $\varepsilon$}{epsilon}-factorised differential equations}
\label{sec:EpsilonBasis}

We employ integration-by-parts (IBP) reduction~\cite{Chetyrkin:1981qh} with Laporta's algorithm~\cite{Laporta:2000dsw} to reduce all integrals in the PB and HB families to a basis of master integrals. 
In particular, we use \texttt{Reduze~2}~\cite{vonManteuffel:2012np} and \texttt{Finred}, the latter being a private implementation employing finite-field methods~\cite{vonManteuffel:2014ixa, Peraro:2016wsq}. 
For the penta-box family, we find $111$ master integrals.
To this end, we also include a non-IBP relation that can be derived by examining Symanzik polynomials using \texttt{Reduze~2} and relates a two-point to a one-point subtopology.
For the hexa-box family, we find $121$ master integrals, $66$ of which are not shared with the penta-box family. 
Therefore, the total number of different master integrals is $177$.
Furthermore, as can be seen from Figure~\ref{fig:Families}, several subtopologies of the two families are related to each other by permutations of external massless momenta, such that the number of integrals posing actual challenges is significantly smaller than $177$.

High-precision evaluations of Feynman integrals are most commonly performed by solving the differential equations they satisfy~\cite{Kotikov:1990kg}. 
In this work, we derive improved bases for the topologies under consideration, such that the differential equations take on a particularly convenient $\varepsilon$-factorised form~\cite{Henn:2013pwa}:
\begin{align}
\label{eq:MI:Deq}
    \mathrm{d}\vec{I} = \varepsilon \, A \, \vec{I} = \varepsilon\sum_k \alpha_k \, A_k \, \vec{I} \, , 
\end{align}
where $\vec{I}$ is the vector of master integrals, and the $\varepsilon$-independent matrix $A \equiv \sum_k \alpha_k \, A_k$ is written as a linear combination of independent one-forms $\alpha_k$, called letters, times $\mathbb{Q}$-valued matrices $A_k$. 
Given a differential equation of the form eq.~\eqref{eq:MI:Deq} and boundary values, one may straightforwardly solve it order-by-order in $\varepsilon$ in terms of iterated integrals, providing an efficient means to compute the Laurent series of the basis.

In the following, we describe our basis and strategies employed to find it. 
In some cases we found it useful to generalise basis definitions described in \cite{FebresCordero:2023pww,Abreu:2021smk}.
In this section, we use numerical samples of differential equations throughout and merely verify their $\varepsilon$-factorisation. 
The symbolic form of the equations is later reconstructed by employing a suitable ansatz, see Sections~\ref{sec:Alphabet}~and~\ref{sec:DeqsPermutations}.

\subsection{Maximal cut}

To find the basis, we work bottom-up in the number of propagators and for each sector start the analysis on the maximal cut, where all propagators are put on-shell and only irreducible scalar products remain as integration variables. 
For many sectors, we use integrand analysis at $\varepsilon = 0$, employing the package \texttt{DlogBasis}~\cite{Henn:2020lye}. 
To this end, we work in loop-by-loop Baikov representations~\cite{Baikov:1996iu, Frellesvig:2017aai, Harley:2017qut, Frellesvig:2024ymq}. 
We find that integrals with constant leading singularities at $\varepsilon = 0$ are often good candidates for canonical master integrals on the maximal cut.

On the other hand, it is well-known that such an analysis at $\varepsilon = 0$ may not always be sufficient to find sufficiently many canonical masters for processes with five external particles~\cite{Chicherin:2018old}. 
Instead, several of the five-point masters employed here can be most conveniently constructed using Gram determinant numerators
\begin{align}
    G_{ij}^\mathrm{fam} \equiv G(\{k_i,p_2,p_3,p_4,p_5\}, \{k_j,p_2,p_3,p_4,p_5\}) \, .
\end{align}
The explicit expression of $G_{ij}^\mathrm{fam}$ in terms of Mandelstam invariants and inverse propagators depends on the family $\mathrm{fam} \in \{\mathrm{PB},\mathrm{HB}\}$.
It is easy to see that all $G_{ij}^\mathrm{fam}$ vanish if $\varepsilon = 0$, and often, employing a Baikov representation, multiplication with such an object can be related to a dimension shift (of sub-loops). 
In our case, all integrals with numerators of this kind could be adapted from refs.~\cite{FebresCordero:2023pww, Abreu:2021smk}.

For other subsectors, we start out by searching for a suitable basis of master integrals such that the differential equations are linear in $\varepsilon$. 
This involves a trial-and-error procedure. 
As usual, we find that for low numbers of propagators, such integrals can often be found by squaring propagators, while for higher numbers of propagators, integrals with irreducible numerators are more appropriate. 
Then, we may systematically ``clean up'' the differential equations on the maximal cut (see e.g.\ ref.~\cite{Gehrmann:2014bfa}). 
Indeed, it is easy to see that if a basis $\vec{I}^\prime$ satisfies differential equations linear in $\varepsilon$ and $W$ is the Wronski matrix for the $\varepsilon^0$-term
\begin{align}
    \mathrm{d}\vec{I}^\prime = (B_0+\varepsilon\, B_1)\,\vec{I}^\prime \,, \qquad 
    \mathrm{d}W = B_0 \, W \, ,
\end{align}
then
\begin{align}
    \vec{I} \equiv W^{-1} \, \vec{I}^\prime
\end{align}
satisfies $\varepsilon$-factorised differential equations. 
The matrix $W$ can be found by explicit solution of its differential equation, or by analysing contour integrals around poles in the integration variables that remain on the maximal cut~\cite{Primo:2016ebd}.

For sectors with massless box-subloops, we also employ the method of building blocks (see e.g.\ ref.~\cite{Dlapa:2022nct}), where one uses one-loop canonical integrals to build two-loop candidates. 
Namely, we may choose numerators which normalise the leading singularities of the one-loop box subdiagram to a constant and then often obtain good candidates also for the considered two-loop integrals.

As an example, we give the bases for the top sectors, each encompassing three master integrals,
\begin{align}
    \IPB_{109} & = \varepsilon^4 \, r_2 \, s_{34} \, 
        \IPB_{11111111} [D_9^\mathrm{PB}] \, , \nonumber\\
    \IPB_{110} & = \frac{32 \, \varepsilon^4 \, s_{34}}{r_1} \, 
        \IPB_{11111111} [G_{12}^\mathrm{PB}] \, , \nonumber\\
    \IPB_{111} & = \frac{32 \, \varepsilon^4 \, s_{34}}{r_1} \, 
        \IPB_{11111111} [G_{11}^\mathrm{PB}] \, , \nonumber\\
    \IHB_{64} & = \varepsilon^4 \, r_2 \, 
        \IHB_{11111111} [D_9^\mathrm{HB} \, (D_3^\mathrm{HB} + 
        D_4^\mathrm{HB} - D_9^\mathrm{HB} - s_{34}) - D_3^\mathrm{HB} \, D_4^\mathrm{HB}] \, , \nonumber\\
    \IHB_{65} & = \frac{32 \, \varepsilon^4}{r_1} \, 
        \IHB_{11111111} [G_{11}^\mathrm{HB} \, D_9^\mathrm{HB}] \, , \nonumber\\
    \IHB_{66} & = \frac{32 \, \varepsilon^4}{r_1} \, 
        \IHB_{11111111} [G_{11}^\mathrm{HB} \, 
        (D_3^\mathrm{HB} + D_4^\mathrm{HB} - D_9^\mathrm{HB} - s_{34})] \, .
\label{eq:MI:TopSector}
\end{align}
Here, $\IPB_{k = 1\dots 111}$ and $\IHB_{k = 1\dots 66}$ denote the canonical master integrals, numerators of integrals are indicated in square brackets and for this reason we also drop the last three propagator indices. 
The functions $r_1$ and $r_2$ are square roots, satisfying
\begin{align}
\label{eq:Roots:12}
    r_1^2 & = 16 \, \Delta_5 \, , \nonumber\\
    r_2^2 & = m^4 s_{12}^2 + 2 \, m^4 s_{12} s_{15} - 2 \, m^2 s_{12}^2 s_{15} + m^4 s_{15}^2  - 2 \, m^2 s_{12} s_{15}^2 + s_{12}^2 s_{15}^2 - 4 \, m^4 s_1 s_{34} \nonumber\\
        & \hspace{11pt} + 2 \, m^2 s_1 s_{12} s_{34} + 2 \, m^2 s_1 s_{15} s_{34} + 4 \, m^2 s_{12} s_{15} s_{34} - 2 \, s_1 s_{12} s_{15} s_{34} - 4 \, m^2 s_1 s_{34}^2 \nonumber\\
        & \hspace{11pt} + s_1^2 s_{34}^2 \, .
\end{align}
Note that although $r_1^2 = 16 \, \Delta_5 = \mathrm{tr}_5^2$, we do not identify $r_1$ with $\mathrm{tr}_5$ because we desire to have parity even master integrals, whereas the latter is parity odd.

\begin{figure}[t]
\centering
\begin{tabular}{llll}
    \raisebox{2pt}{\includegraphics[scale = 0.75]{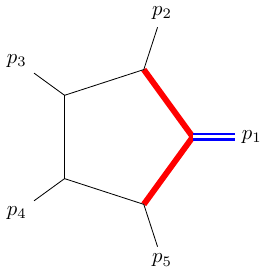}} & 
    \includegraphics[scale = 0.75]{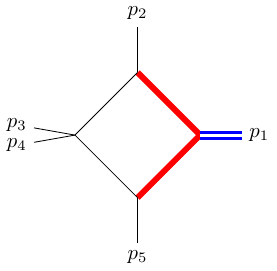} & 
    \raisebox{8pt}{\includegraphics[scale = 0.75]{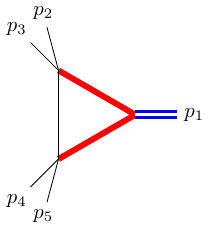}} & 
    \raisebox{33pt}{\includegraphics[scale = 0.75]{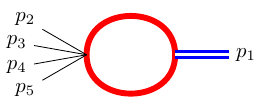}}
\end{tabular}
\caption{One-loop Feynman integral topologies associated with the square roots $r_{1\dots 4}$.}
\label{fig:R1234}
\end{figure}

\begin{figure}[t]
\centering
\begin{tabular}{ccc}
    \raisebox{15pt}{\includegraphics[scale = 0.75]{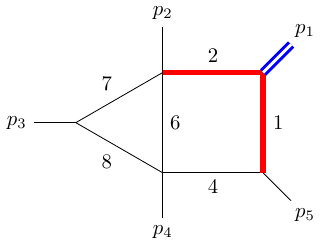}} & 
    \raisebox{15pt}{\includegraphics[scale = 0.75]{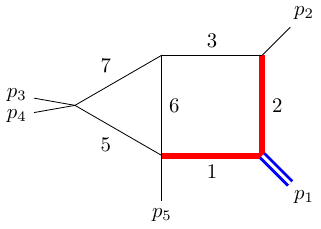}} & 
    \includegraphics[scale = 0.75]{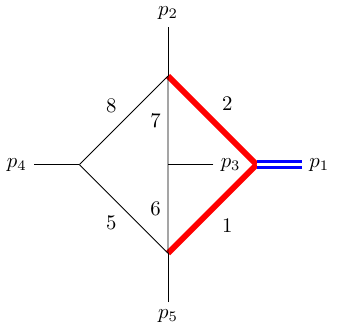} 
\end{tabular}
\caption{Two-loop Feynman integral topologies associated with the square roots $r_{5\dots 7}$. 
The first two diagrams are subtopologies of the penta-box, the third one of the hexa-box.}
\label{fig:R567}
\end{figure}

For canonical bases of subsectors, we also need five other types of square roots $r_{i = 3\dots 7}$, defined to satisfy
\begin{align}
    r_3^2 & = s_1^2 + s_{23}^2 + s_{45}^2 - 2 \, s_1 s_{23} - 2 \, s_1 s_{45} - 2 \, s_{23} s_{45} \, , \nonumber\\
    r_4^2 & = - s_1 \, (4 \, m^2 - s_1) \, , \nonumber\\
    r_5^2 & = m^4 s_{12}^2 - 2 \, m^2 s_{12}^2 s_{15} + s_{12}^2 s_{15}^2 + 2 \, m^4 s_{12} s_{23} - 2 \, m^2 s_{12} s_{15} s_{23} + m^4 s_{23}^2 \nonumber\\
        & \hspace{11pt} + 2 \, m^2 s_1 s_{12} s_{34} + 4 \, m^2 s_{12} s_{15} s_{34} - 2 \, s_1 s_{12} s_{15} s_{34} - 2 \, m^2 s_1 s_{23} s_{34} + 4 \, m^2 s_{15} s_{23} s_{34} \nonumber\\
        & \hspace{11pt} - 4 \, m^2 s_1 s_{34}^2 + s_1^2 s_{34}^2 - 2 \, m^4 s_{12} s_{45} + 4 \, m^2 s_{12} s_{15} s_{45} - 2 \, s_{12} s_{15}^2 s_{45} - 2 \, m^4 s_{23} s_{45}  \nonumber\\
        & \hspace{11pt} + 2 \, m^2 s_{15} s_{23} s_{45} - 2 \, m^2 s_1 s_{34} s_{45} - 4 \, m^2 s_{15} s_{34} s_{45} + 2 \, s_1 s_{15} s_{34} s_{45} + m^4 s_{45}^2 \nonumber\\
        & \hspace{11pt} - 2 \, m^2 s_{15} s_{45}^2 + s_{15}^2 s_{45}^2 \, , \nonumber\\
    r_6^2 & = m^4 s_{12}^2 + 2 \, m^4 s_{12} s_{15} - 2 \, m^2 s_{12}^2 s_{15} + m^4 s_{15}^2 - 2 \, m^2 s_{12} s_{15}^2 + s_{12}^2 s_{15}^2 - 4 \, m^4 s_1 s_{34} \nonumber\\
        & \hspace{11pt} + 4 \, m^2 s_1 s_{12} s_{34} - 2 \, m^2 s_{12}^2 s_{34} + 2 \, m^2 s_{12} s_{15} s_{34} - 2 \, s_{12}^2 s_{15} s_{34} + s_{12}^2 s_{34}^2 \, , \nonumber\\
    r_7^2 & = s_{12}^2 s_{15}^2 - 2 \, s_{12}^2 s_{15} s_{23} + s_{12}^2 s_{23}^2 - 4 \, m^2 s_{12} s_{15} s_{34} + 4 \, m^2 s_{12} s_{23} s_{34} - 4 \, m^2 s_{15} s_{23} s_{34} \nonumber\\
        & \hspace{11pt} + 2 \, s_{12} s_{15} s_{23} s_{34} + 4 \, m^2 s_{23}^2 s_{34} - 2 \, s_{12} s_{23}^2 s_{34} + s_{23}^2 s_{34}^2 - 2 \, s_{12} s_{15}^2 s_{45} \nonumber\\
        & \hspace{11pt} + 2 \, s_{12} s_{15} s_{23} s_{45} - 4 \, m^2 s_{12} s_{34} s_{45} + 4 \, m^2 s_{15} s_{34} s_{45} + 2 \, s_{12} s_{15} s_{34} s_{45} \nonumber\\
        & \hspace{11pt} - 8 \, m^2 s_{23} s_{34} s_{45} + 2 \, s_{12} s_{23} s_{34} s_{45} + 2 \, s_{15} s_{23} s_{34} s_{45} - 2 \, s_{23} s_{34}^2 s_{45} + s_{15}^2 s_{45}^2 \nonumber\\
        & \hspace{11pt}  + 4 \, m^2 s_{34} s_{45}^2 - 2 \, s_{15} s_{34} s_{45}^2 + s_{34}^2 s_{45}^2 \, .
\label{eq:Roots:34567}
\end{align}
The roots $r_{1\dots 4}$ are already known from the one-loop integrals~\cite{Asteriadis:2026olo}, $r_{5\dots 6}$ arise first in penta-box subtopologies and $r_7$ originates from a subtopology of the hexa-box. 
We display diagrams for integrals whose leading singularities involve these square roots in Figures~\ref{fig:R1234} and~\ref{fig:R567}. 
Additionally, already to define canonical integrals for the uncrossed topology, we need crossed versions of $r_3$, $r_5$ and $r_6$ with permutations of massless momenta.

\begin{figure}[t]
\centering
\includegraphics[scale = 0.75]{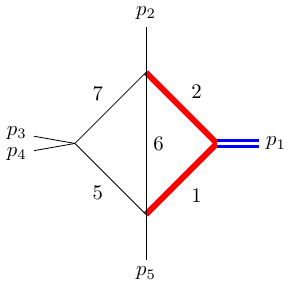}
\caption{Subtopology of the penta-box involving a nested square root.}
\label{fig:NestedRoots}
\end{figure}

However, the penta-box also contains a four-point subtopology, depicted in Figure~\ref{fig:NestedRoots}, whose differential equations can not be transformed to $\varepsilon$-factorised form using rational functions and simple square roots only. 
Instead, this topology requires the introduction of nested square roots $\nr_{1a/b}$, defined to satisfy
\begin{align}
    \nr_{1a}^2 & = P_b + \nr_{1i} \, , \nonumber\\
    \nr_{1b}^2 & = P_b - \nr_{1i} \, , \nonumber\\
    \textnormal{where}\quad \nr_{1i}^2 & = P_b^2 - P_c \, , \nonumber\\
    P_b & = s_1 \, ((s_{12} - s_{34})^2 + (s_{15} - s_{34})^2) 
        - 2 \, m^2 \, ((s_{12} + s_{15})^2 - 4 \, s_1 s_{34}) \, , \nonumber\\
    P_c & = - s_1 \, (4 \, m^2 - s_1) \, (s_{12} - s_{15})^2 \, 
        (s_{12} + s_{15} - 2 \, s_{34})^2 \, .
\label{eq:NR}
\end{align}
Both $\nr_{1a}$ and $\nr_{1b}$ always appear together with a square root $\sqrt{s_1}$. 
Nevertheless, we prefer to keep the latter separate to simplify the application of the analytic continuation algorithm described in Section~\ref{sec:FunctionBasis} and Appendix~\ref{app:AnalyticContinuation}.
Moreover, we note that $\nr_{1a}$ and $\nr_{1b}$ are actually not multiplicatively independent from the root $r_4$; up to branches, they satisfy the relation
\begin{align}
\label{eq:NR:Relation}
    \nr_{1a} \, \nr_{1b} = r_4 \, (s_{12} - s_{15}) \, (s_{12} + s_{15} - 2 \, s_{34}) \, .
\end{align}
For this reason, following ref.~\cite{FebresCordero:2023pww}, one could eliminate $\nr_{1b}$ altogether and express it instead as $r_4 \, (s_{12} - s_{15}) \, (s_{12} + s_{15} - 2 \, s_{34})\,\nr_{1a}^{-1}$. 
This, however, introduces more complicated expressions into the letters. Therefore, we prefer to work with both $\nr_{1a}$ and $\nr_{1b}$.

Fortunately, by inspection we see that Figure~\ref{fig:NestedRoots} is very similar to the \textit{kite}${}_7$ subtopology studied in ref.~\cite{FebresCordero:2023pww}, and thus we may guess a suitable basis for this sector by adaptation from the reference. 
We find seven master integrals, given by
\begin{align}
    \IPB_{71} & = \varepsilon^4 \, r_3^{(3)} \, \IPB_{11001110} \, , \nonumber\\
    \IPB_{72} & = - \varepsilon^3 \, r_2 \, \IPB_{11001210} \, , \nonumber\\
    \IPB_{73} & = - \varepsilon^3 \, (s_1 - s_{15}) \, s_{34} \, \IPB_{11002110} \, , \nonumber\\
    \IPB_{74} & = - \varepsilon^3 \, (s_1 - s_{12}) \, s_{34} \, \IPB_{11001120} \, , \nonumber\\
    \IPB_{75} & = \frac{1}{2 \, ( - 2 \, m^2 s_1 + s_1 s_{12} - s_{12}^2 + s_1 s_{15} - s_{15}^2)}
        \big(Q_1 \, \varepsilon^2 \, (\IPB_{12002110} + \IPB_{21001120})\nonumber\\
        & \hspace{11pt} + Q_2 \, \varepsilon^3\IPB_{11001120} + Q_3 \, \varepsilon^3\IPB_{11002110} + 
        Q_4 \, \varepsilon^3 \, \IPB_{12001110} + Q_5 \, \varepsilon^3 \, \IPB_{21001110}\nonumber\\
        & \hspace{11pt} + Q_6 \, \IPB_{11001210} + \textnormal{(subsectors)}\big) \, , \nonumber\\
    \IPB_{76} & = \varepsilon^3 \, \sqrt{s_1} \, 
        \bigg(\nr_{1b} + \frac{r_4 \, \nr_{1a}}{s_1}\bigg)\IPB_{21001110} + 
        \varepsilon^3 \, \sqrt{s_1} \, 
        \bigg( - \nr_{1b} + \frac{r_4 \, \nr_{1a}}{s_1}\bigg)\IPB_{12001110} \, , \nonumber\\
    \IPB_{77} & = \varepsilon^3 \, \sqrt{s_1} \, 
        \bigg(\nr_{1a} + \frac{r_4 \, \nr_{1b}}{s_1}\bigg)\IPB_{21001110} + 
        \varepsilon^3 \, \sqrt{s_1} \, 
        \bigg( - \nr_{1a} + \frac{r_4 \, \nr_{1b}}{s_1}\bigg)\IPB_{12001110} \, , 
\label{eq:MI:NR}
\end{align}
where the root $r_3^{(3)}$ is a permutation of $r_3$ satisfying $(r_3^{(3)})^2 \equiv (s_{12} + s_{15})^2 - 4 \, s_1 s_{34}$, and $Q_i$ are polynomials in kinematic invariants whose explicit forms will not be relevant in the following.

\subsection{Subtopology contributions}
For each sector, after having derived a master integral basis satisfying $\varepsilon$-factorised differential equations on the maximal cut, we relax the cut conditions to include also subsector contributions. 
Sufficiently often, we find a dependency of the differential equations linear in $\varepsilon$. In such a case, the differential equations can be systematically cleaned up by ``correcting'' a master integral with subtopology integrals, see e.g.\ ref.~\cite{Gehrmann:2014bfa}. 
We note that this requires the knowledge of the functional form of the differential equations at $\varepsilon = 0$, contrary to the statement at the beginning of this section that we would like to work with numerical samples only. 
Thus, we need to perform functional reconstruction, which is in the present case easy enough to be solved by guessing of denominators from partly-symbolic samples and afterwards determining numerators from interpolating multiple samples. 
Moreover, we use ansatzing strategies similar to the ones described in ref.~\cite{Abreu:2024fei}. 
In particular, assuming that an inhomogeneous term in the differential equation at $\varepsilon = 0$ can be integrated out with an algebraic function times a canonical subsector integral, it is sufficient for the clean-up to fit only coefficients in the derivatives of an ansatz function instead of reconstructing the partial derivatives directly.

For integrals involving at most simple square roots, clean-up strategies have been described in many references. 
In the case of a differential equation linear in $\varepsilon$, they amount to integrating out its $\varepsilon^0$-terms for each entry in the matrix separately (see e.g.\ ref.~\cite{Gehrmann:2014bfa}).
For cases with nested square roots, the differential equations occurring in the clean-up step are again most easily solved by differentiating an ansatz.

Consider an integral $I^\prime$ not involving nested roots that has $\varepsilon$-factorised differential equations on the maximal cut, but couples at $\varepsilon = 0$ to a subsector involving nested square roots:
\begin{align}
    \mathrm{d}I^\prime|_{\varepsilon = 0} = \vec{a}\cdot\vec{J}^\mathrm{alm}
    + \textnormal{(terms independent of $I^\prime$ and $\vec{J}^\mathrm{alm}$)} \, ,
\end{align}
where $\vec{a}$ is a $2$-vector of algebraic functions involving at most single square roots and $\vec{J}^\mathrm{alm} \equiv (I_{76}^\mathrm{alm},I_{77}^\mathrm{alm})^T$ is a vector of almost-canonical subsector master integrals whose transformation to canonical form involves nested square roots. 
Thus,
\begin{align}
    \mathrm{d}\vec{J}^\mathrm{alm}|_{\varepsilon = 0} = M \, \vec{J}^\mathrm{alm} \, , 
\end{align}
where the Wronski matrix $W$ satisfying $\mathrm{d}W = M \, W$ contains nested square roots, and $\vec{J} \equiv W^{-1} \, \vec{J}^\mathrm{alm}$ satisfies $\varepsilon$-factorised differential equations. 
The ansatz for the canonical master integral $I$ satisfying $\varepsilon$-factorised differential equations also when projected onto the space spanned by $\vec{J}^\mathrm{alm}$ is
\begin{align}
    I = I^\prime + \vec{v}\cdot\vec{J}^\mathrm{alm} \, , 
\end{align}
for a vector $\vec{v}$ of algebraic functions involving at most single square roots to be determined in the following. 
Assembling the above, it follows that $\vec{v}$ needs to satisfy
\begin{align}
\label{eq:CleanUp:Deq}
    (\mathrm{d}\vec{v})\cdot\vec{J}^\mathrm{alm} + \vec{v}\cdot M \, \vec{J}^\mathrm{alm}
    + \vec{a}\cdot\vec{J}^\mathrm{alm} = 0 \, .
\end{align}
Since the $I_{76}^\mathrm{alm}$- and $I_{77}^\mathrm{alm}$-components of eq.~\eqref{eq:CleanUp:Deq} are a system of two linear differential equations for the two entries of $\vec{v}$, determining the solution for $\vec{v}$ is in principle straightforward. 
However, the usual method by variation of constants proceeds by first determining the solution to the corresponding homogeneous system, which is given by rows of $W^{-1}$ and hence involves nested square roots again. 
Since the solution for the inhomogeneous differential equation contains at most single square roots, nested ones need to cancel out in the solution process, but complicate intermediate algebraic steps and hinder computer algebra systems to determine the solution directly. 
Hence, it is more efficient to solve eq.~\eqref{eq:CleanUp:Deq} by differentiating a suitable ansatz for $\vec{v}$.

In an analogous fashion, we may clean up subtopology contributions for a topology that involves a nested square root on the maximal cut.
We assume that we already know that two master integrals $\vec{I}^\prime$ can be rotated to canonical form on their maximal cut using a matrix with nested square roots, but their differential equations have not yet $\varepsilon$-factorised dependency on a canonical subtopology integral $J$. 
Thus,
\begin{align}
    \mathrm{d}\vec{I}^\prime|_{\varepsilon = 0} = M \, \vec{I}^\prime + \vec{a} \, J + \textnormal{(terms independent of $\vec{I}^\prime$ and $J$)} \, , 
\end{align}
where the Wronski matrix satisfying $\mathrm{d}W = M \, W$ involves nested square roots and $\vec{a}$ is a vector of algebraic functions. 
In this case, we want to improve $\vec{I}^\prime$ to almost-canonical masters $\vec{I}^\mathrm{alm}$ such that $\vec{I} \equiv W^{ - 1} \, \vec{I}^\mathrm{alm}$ satisfies $\varepsilon$-factorised differential equations on the maximal cut as well as when projected onto $J$. 
As ansatz we use
\begin{align}
    \vec{I}^\mathrm{alm} = \vec{I}^\prime + \vec{v} \, J \, , 
\end{align}
such that
\begin{align}
    \mathrm{d}\vec{I}^\mathrm{alm}|_{\varepsilon = 0} = M \, \vec{I}^\mathrm{alm} - M \, \vec{v} \, J + (\mathrm{d}\vec{v}) \, J + \vec{a} \, J + \textnormal{(terms independent of $\vec{I}^\mathrm{alm}$ and $J$)} \, , 
\end{align}
and thus $\vec{v}$ needs to satisfy
\begin{align}
    (\mathrm{d}\vec{v}) \, J - M \, \vec{v} \, J + \vec{a} \, J = 0 \, .
\end{align}
This can also be most easily solved with an ansatz, since $\vec{v}$ does not contain nested square roots while the solution to the homogeneous differential equation would do. 
Using such ansätze, we obtain master integrals satisfying $\varepsilon$-factorised differential equations also beyond the maximal cut.

In the ancillary files, we provide an almost-canonical basis consisting of $111+66$ integrals for the integral families in \texttt{Reduze~2} format, which differs from the final basis only by multiplication with matrices containing the square-root contributions.
Standard IBP reduction programs can not deal with multi-valued objects, and for this reason we give these matrices separately in \texttt{Mathematica} format.

\section{Alphabet}
\label{sec:Alphabet}
In the previous section, we discussed how to find good basis integrals for the penta-box and hexa-box topologies, where the $\varepsilon$-factorisation of the differential equations has merely been verified at numerical phase-space points. 
As a next step, to determine the symbolic form of eq.~\eqref{eq:MI:Deq}, we need to obtain the letters $\alpha_k$.

Let us consider the canonical case, where the one-forms $\alpha_k$ can be written as
\begin{align}
    \alpha_k = \mathrm{d}\log(W_k)
\end{align}
for algebraic functions $W_k(\vec{x})$. 
We note that in the case of nested square roots, it is so far unclear whether a canonical form exists~\cite{FebresCordero:2023pww}, but for the penta-box and hexa-box discussed in this work, we will demonstrate that it is indeed possible to find one.

\subsection{Letters without nested square roots}
\label{subsec:Alphabet:WithoutNested}

First, we focus on the letters which are independent of nested square roots. 
Several of these are already known from the one-loop integrals~\cite{Asteriadis:2026olo}, others have to be newly determined.

We need to find rational functions in $\vec{x}$ and square roots $r_{1\dots 7}$ such that their $\mathrm{d}\log$'s describe the full singularity structure of the integral families. 
To simplify the problem, we first note that square roots may not appear at arbitrary places in the differential equations. 
It is well-known that the signs of square roots in the canonical basis are arbitrary and need to drop out of predictions for scalar integrals. 
Algebraically, the two choices of square root are equivalent. 
Consequently, they are related by a so-called Galois transformation (which for simple square roots amounts to merely changing signs), and we may study how the master integrals behave under such transformations. 
Obviously, integrals without square-root prefactors are even, i.e.\ invariant, while integrals with a square-root prefactor also change sign under the transformation, thus they are called odd with respect to the sign change of this square root. 
In more mathematical terms, the Galois group of a simple square root is isomorphic to $\mathbb{Z}/2\mathbb{Z}$. 
Even integrals transform in the trivial representation of this group, whereas odd integrals transform in the sign representation.

Now, the differential equation~\eqref{eq:MI:Deq} couples integrals with different Galois transformation behaviours to each other. 
Consequently, the entries of the matrix $A$ appearing in it can not take arbitrary forms, but must also transform appropriately such that both sides of the differential equation behave in the same way. 
As an example, let us consider a canonical integral $I_i = r \, I_i^\mathrm{alm}$, where $I_i^\mathrm{alm}$ is a Galois-invariant linear combination of scalar integrals with rational function coefficients and $r$ is a square root. 
The differential equation for $I_i$ takes on the form
\begin{align}
    \mathrm{d}I_i = \varepsilon\sum_j A_{ij} \, I_j \, .
\end{align}
Here, the left-hand side changes sign under $r\rightarrow -r$, and thus also the right-hand side must behave in this way. 
Therefore, we see that for all $I_j$ that are invariant under the sign change of $r$, the corresponding entry $A_{ij}$ must change sign. 
Contrarily, for $I_j$ which also carry an $r$-prefactor and no other algebraic prefactor (in particular $I_i$ itself), the entry $A_{ij}$ must be invariant under $r\rightarrow - r$. 
Finally, if $I_j = r^\prime \, I_j^\mathrm{alm}$ depends on another square root $r^\prime$ but not on $r$, then the entry $A_{ij}$ must change sign both under $r\rightarrow -r$ and under $r^\prime\rightarrow -r^\prime$.
In the present work, there occurred no integrals depending on two roots simultaneously and thus we do not need to consider this case.
We can make these transformation behaviours manifest by our choice of letters, namely we choose
\begin{align}
\label{eq:Ansatz:Aij}
    A_{ij} = \begin{cases} 
    \sum_k a_{ijk} \, \mathrm{d}\log\Big(\frac{q_k + s_k \, r}{q_k - s_k \, r}\Big) \, , & 
        \textnormal{if $I_j$ is even},\\[4pt]
    a_{ij0} \, \mathrm{d}\log(r) + \sum_k a_{ijk} \, \mathrm{d}\log(q_k) \, , & 
        \textnormal{if $I_j$ is odd with respect to $r\rightarrow - r$},\\[4pt]
    \sum_k a_{ijk} \, \mathrm{d}\log\Big(\frac{q_k + s_k \, r \, r^\prime}{q_k - s_k \, r \, r^\prime}\Big) \, , & 
        \textnormal{if $I_j$ is odd with respect to $r^\prime\rightarrow - r^\prime$},
    \end{cases}
\end{align}
where $a_{ijk}$ are rational numbers and $q_k$, $s_k$ are polynomials in kinematic invariants. 
Note that $\mathrm{d}\log(r) = \mathrm{d}\log(r^2)/2$ is invariant under $r\rightarrow -r$, which is why it is included in the second case of eq.~\eqref{eq:Ansatz:Aij}.
Thus, we see that all letters which are independent of nested square roots may be chosen to have one of three forms:
\begin{align}
\label{eq:Ansatz:Polylog}
    \alpha_k = \begin{cases}\mathrm{d}\log(q_k) \ \textnormal{or}\ \mathrm{d}\log(r_i)\, , & 
        \textnormal{for even letters}, \\[4pt]
    \mathrm{d}\log\Big(\frac{q_k + s_k \, r_i}{q_k - s_k \, r_i}\Big) \, , & 
        \textnormal{for single square-root-odd letters}, \\[4pt]
    \mathrm{d}\log\Big(\frac{q_k + s_k \, r_i \, r_j}{q_k - s_k \, r_i \, r_j}\Big) \, , & 
        \textnormal{for double square-root-odd letters}.
    \end{cases}
\end{align}

In this work, even letters are mostly determined using a combination of the codes \texttt{Baikovletter}~\cite{Jiang:2024eaj} and \texttt{SOFIA}~\cite{Correia:2025wtb}. 
For one letter arising in a sub-topology of the hexa-box, we rely on symbolic IBP reduction and derivation of differential equations on the maximal cut.

Once the even alphabet is known, square-root-odd letters can be found in an algorithmic way. 
We use again \texttt{Baikovletter}, as well as a private implementation of the algorithm described in refs.~\cite{Heller:2019gkq, Matijasic:2024too}. 
In brief, this strategy is based on the observation that all singularities of the Feynman integrals are often already captured by even letters. 
Hence, for a single square-root-odd-letter as in eq.~\eqref{eq:Ansatz:Polylog}, one conjectures the factorisation (see e.g.\ ref.~\cite{Heller:2019gkq})
\begin{align}
\label{eq:FactorisationConjecture}
    (q_k + s_k \, r_i) \, (q_k - s_k \, r_i) = q_k^2 - s_k^2 \, r_i^2 \overset{!}{=} 
    c \times \textnormal{(product of even letters)},
\end{align}
where $c$ is just a constant (empirically one finds $c \in \{\pm 4\}$). 
This may be used to determine square-root-odd letters by guessing $s_k$, constructing power products
\begin {align}
    g_k \equiv c \times \textnormal{(product of even letters)}
\end{align}
and then testing whether $g_k + s_k^2 \, r_i^2$ becomes a perfect square. 
If it is, we obtain a single square-root-odd letter candidate by setting
\begin{align}
    q_k \equiv \sqrt{g_k + s_k^2 \, r_i^2} \,.
\end{align}
We proceed analogously for double square-root-odd letters.

Alternatively, we may employ rationalisations of square roots to determine odd letters. 
For example, for letters involving the square root $r_7$, we can determine a variable transformation $\vec{x} = \vec{x}(\vec{y})$ rationalising this root, e.g.\ using the package \texttt{RationalizeRoots}~\cite{Besier:2019kco}. 
We rewrite all $\mathrm{d}\log$ arguments of even letters in terms of the variables $\vec{y}$, in terms of which they are rational functions, and factorise them. 
Any irreducible polynomial factor occurring in such a factorisation is then transformed back to the original variables $\vec{x}$, re-introducing the square root $r_7$. 
Then, $\mathrm{d}\log$'s of the resulting algebraic functions are candidates for letters involving $r_7$, however, they take on the form $\mathrm{d}\log(q_k + s_k \, r_7)$ instead of eq.~\eqref{eq:Ansatz:Polylog}. 
We replace them by $\mathrm{d}\log((q_k + s_k \, r_7)/(q_k - s_k \, r_7))$ to obtain candidates with desirable behaviour under Galois transformations. 
Due to the factorisation~\eqref{eq:FactorisationConjecture}, such a replacement does not change the span of the alphabet.

\subsection{Letters with nested square roots: general considerations}
The derivation of letters involving nested square roots is more intricate, and we use different strategies to find them. 
Firstly, as noticed in ref.~\cite{Becchetti:2025oyb}, to simplify the problem we may again use Galois theory. 
For nested square roots, instead of the even/odd decomposition discussed in the previous section, we find a different structure. 
Indeed, the Galois group of a nested square root is isomorphic to $(\mathbb{Z}/2\mathbb{Z}) \times (\mathbb{Z}/2\mathbb{Z})$, whose action is determined by
\begin{align}
    \pi_{(1,0)} \begin{pmatrix} \nr_{1a} \\ \nr_{1b} \end{pmatrix} & = 
        - \begin{pmatrix} \nr_{1a} \\ \nr_{1b} \end{pmatrix} \, , \nonumber\\
    \pi_{(0,1)} \begin{pmatrix} \nr_{1a} \\ \nr_{1b} \end{pmatrix} & = 
        \begin{pmatrix} \nr_{1b} \\ \nr_{1a} \end{pmatrix} = 
        \sigma_1\begin{pmatrix} \nr_{1a} \\ \nr_{1b} \end{pmatrix} \, .
\end{align}
Here, $\pi_{(1,0)}$ and $\pi_{(0,1)}$ on the left-hand side denote the actions of the group elements $(1,0)$ and $(0,1)$, respectively, and $\sigma_1$ is the first Pauli matrix.

For this reason, we find a duplet structure on the master integrals depending on nested square roots. 
For example, one such duplet is given by the master integrals $(\IPB_{76},\IPB_{77})^T$ defined in eq.~\eqref{eq:MI:NR}. 
Under the action of the element $(1,0)$ of the Galois group, this is transformed into $ - (\IPB_{76},\IPB_{77})^T$, and under the action of $(0,1)$ into $(\IPB_{77},\IPB_{76})^T = \sigma_1 \, (\IPB_{76},\IPB_{77})^T$.

We would like to choose the alphabet with a similar structure, such that we can argue analogously to the previous section that certain letters may only appear at certain points in the differential equation (see ref.~\cite{Becchetti:2025oyb}). 
For this reason, we choose the letters involving nested square roots in duplets $(\alpha_i,\alpha_{i + 1})^T$, where we write $\alpha_i$ only using $\nr_{1a}$ but not $\nr_{1b}$, and $\alpha_{i + 1} = \alpha_i|_{\nr_{1i}\rightarrow - \nr_{1i}}$ (we recall that $\nr_{1i}\rightarrow - \nr_{1i}$ implies $\nr_{1a}\leftrightarrow \nr_{1b}$). 
A general ansatz for a duplet of such letters, if they can be written in $\mathrm{d}\log$-form, is
\begin{align}
\label{eq:Letters:Duplet}
    \begin{pmatrix}\alpha_i \\ \alpha_{i+1}\end{pmatrix} = \bigg(\mathrm{d}\log\bigg(\frac{q + s \, \nr_{1i} + t \, \nr_{1a}}{q + s \, \nr_{1i} - t \, \nr_{1a}}\bigg),\mathrm{d}\log\bigg(\frac{q - s \, \nr_{1i} + t \, \nr_{1b}}{q - s \, \nr_{1i} - t \, \nr_{1b}}\bigg)\bigg)^T \, , 
\end{align}
where $q$, $s$ are polynomials in kinematic invariants, and the function $t$ is proportional to $\sqrt{s_1}$ and may or may not involve further square roots.

This representation is not unique, and we can employ its non-uniqueness to derive the letters in a more convenient form (see Subsection~\ref{subsec:Alphabet:WithNested:Derivation}). 
Namely, up to branches, we can rewrite
\begin{align}
    \nr_{1a/b} & = \sqrt{\frac{P_b - (s_{12} - s_{15}) \, (s_{12} + s_{15} - 2 \, s_{34}) \, r_4}{2}} \nonumber\\
    & \hspace{11pt}\pm\sqrt{\frac{P_b + (s_{12} - s_{15}) \, (s_{12} + s_{15} - 2 \, s_{34}) \, r_4}{2}} \equiv \nr^{(a)}\pm\nr^{(b)}.
\label{eq:NR:Rewrite}
\end{align}
Here, $r_4$ is a square root introduced in eq.~\eqref{eq:Roots:34567} and the polynomial $P_b$ has been defined in eq.~\eqref{eq:NR}.
This identity can be proven by simple algebraic manipulations, analogously to the following classical formula for denesting numeric roots going back to Euler (\cite{Euler:2011}, Section IV, Chapter VIII): 
If $a,b \in \mathbb{Q}$ such that $a^2 - b = c^2$ is a perfect square in $\mathbb{Q}$, then
\begin{align}
    \sqrt{a\pm\sqrt{b}} = \sqrt{\frac{a + c}{2}}\pm\sqrt{\frac{a - c}{2}}
\end{align}
(up to branches).

Thus, we may replace the pair $(\nr_{1a},\nr_{1b})$ by the nested square roots $(\nr^{(a)},\nr^{(b)})$. 
Any differential form proportional to $\nr_{1a}$ or $\nr_{1b}$, and so in particular the $\mathrm{d}\log$-forms of eq.~\eqref{eq:Letters:Duplet}, may be rewritten as a linear combination of differential forms proportional to $\nr^{(a)}$ and $\nr^{(b)}$. 
We write this decomposition as
\begin{align}
\label{eq:NR:LetterDecomposition}
    \alpha = \alpha^{(a)} + \alpha^{(b)} \, , 
\end{align}
where $\alpha^{(a)}$ and $\alpha^{(b)}$ are Galois-conjugates of each other and proportional to $\nr^{(a)}$ and $\nr^{(b)}$, respectively. 
As an example, we may derive that up to branches,
\begin{align}
    & \mathrm{d}\log\bigg(\frac{s_1 \, (s_{12} + s_{15} - 2 \, s_{34}) + \sqrt{s_1} \, \nr_{1a}}
    {s_1 \, (s_{12} + s_{15} - 2 \, s_{34}) - \sqrt{s_1} \, \nr_{1a}}\bigg) \nonumber\\
    & \hspace{0.25cm} = \frac{1}{2} \, \mathrm{d}\log\bigg(
    \frac{q + \sqrt{s_1} \, \nr^{(a)} \, (s_1 \, (s_{12} + s_{15} - 2 \, s_{34}) + 
    r_4 \, (s_{12} - s_{15}))}
    {q - \sqrt{s_1} \, \nr^{(a)} \, (s_1 \, (s_{12} + s_{15} - 2 \, s_{34}) + 
    r_4 \, (s_{12} - s_{15}))}\bigg) + 
    (r_4\rightarrow - r_4) \, , 
\label{eq:NR:LetterDecomposition:Example}
\end{align}
where $r_4\rightarrow - r_4$ implies $\nr^{(a)}\leftrightarrow\nr^{(b)}$ and
\begin{align}
    q \equiv 2 \, s_1 \, (m^2 s_{12}^2 - 2 \, m^2 s_{12} s_{15} + s_1 s_{12} s_{15} + m^2 s_{15}^2 - s_1 s_{12} s_{34} - s_1 s_{15} s_{34} + s_1 s_{34}^2) \, .
\end{align}

Since we know from considering the Galois transformation behaviour of the differential equations that all letters must be either proportional to $\nr_{1a}$ or $\nr_{1b}$ or be independent of both, such a decomposition exists for all letters that we need to consider. 
Moreover, eq.~\eqref{eq:NR:Rewrite} may be inverted to give $\nr^{(a)}$ and $\nr^{(b)}$ in terms of $\nr_{1a}$ and $\nr_{1b}$, and thus we see that an expression of the whole system in terms of $\nr^{(a)}$ and $\nr^{(b)}$ is equivalent to an expression in terms of the original $\nr_{1a}$ and $\nr_{1b}$.

Nevertheless, to ensure uniqueness, we choose to express our final results in terms of $\nr_{1a}$ and $\nr_{1b}$ instead of $\nr^{(a)}$ and $\nr^{(b)}$. 
Note that even when using $\nr^{(a)}$ and $\nr^{(b)}$, the square root $\nr_{1i}$ would appear in the differential equations, hence this would not give significant simplifications once the results have been derived.

\subsection{Letters with nested square roots: derivation strategies}
\label{subsec:Alphabet:WithNested:Derivation}

To find letters of the form eq.~\eqref{eq:Letters:Duplet}, we can use different methods. 
On the one hand, we may treat eq.~\eqref{eq:Letters:Duplet} as an ansatz and attempt to generalise the algorithm for the generation of square-root-odd letters from refs.~\cite{Heller:2019gkq, Matijasic:2024too} to this case. 
On the other hand, we can look for sufficiently similar structures obtained by explicit integration of partial derivatives and only later rewrite them in the form of eq.~\eqref{eq:Letters:Duplet}. 
In this work, we use both approaches, together with adaptation of letters from ref.~\cite{FebresCordero:2023pww}.

Firstly, let us consider the algorithm of refs.~\cite{Heller:2019gkq, Matijasic:2024too}. 
To generalise it to the case of nested square roots, we need an analogue of the factorisation conjecture~\eqref{eq:FactorisationConjecture}. 
By inspection, we find that for letters of the form~\eqref{eq:Letters:Duplet}, such an analogue is given by
\begin{align}
    (q + s \, \nr_{1i} + t \, \nr_{1a}) \, 
    (q + s \, \nr_{1i} - t \, \nr_{1a}) \, 
    (q - s \, \nr_{1i} + t \, \nr_{1b}) \, 
    (q - s \, \nr_{1i} - t \, \nr_{1b}) \nonumber\\
    \overset{?}{ = } c \times \textnormal{(product of even letters)} \, , 
\label{eq:FactorisationConjectureNR}
\end{align}
where $c$ is again a constant. 
In the present case, for several letters (in particular if $s = 0$) we find this indeed to be true, with $c \in \{\pm 16\}$. 
To exploit such a factorisation, we first search for polynomials $\tilde{q}$ and $\tilde{s}$ such that
\begin{align}
    \tilde{q}^2 - \tilde{s}^2 \, \nr_{1i}^2 = c \times \textnormal{(product of even letters)} \, , 
\end{align}
using the strategy for the construction of single square-root-odd letters described in Subsection~\ref{subsec:Alphabet:WithoutNested}. 
If we can find such $\tilde{q}$ and $\tilde{s}$, we attempt to determine polynomials $q$, $s$, and $t$ such that
\begin{align}
\label{eq:NR:Letters:qst}
    (q + s \, \nr_{1i} + t \, \nr_{1a}) \, 
    (q + s \, \nr_{1i} - t \, \nr_{1a}) = 
    \tilde{q} + \tilde{s} \, \nr_{1i} \, , 
\end{align}
which are then seen to satisfy eq.~\eqref{eq:FactorisationConjectureNR}.
Eq.~\eqref{eq:NR:Letters:qst} can often be solved with a suitable ansatz, giving us candidates of letters as in eq.~\eqref{eq:Letters:Duplet}.

However, for some letters with $s\neq 0$, this strategy does not produce the desired result.
Investigating these cases more thoroughly, we observe eq.~\eqref{eq:FactorisationConjectureNR} not to hold, rendering the method not applicable. 
Instead, we find a factorisation
\begin{align}
    (q + s \, \nr_{1i} + t \, \nr_{1a}) \, 
    (q + s \, \nr_{1i} - t \, \nr_{1a}) \, 
    (q - s \, \nr_{1i} + t \, \nr_{1b}) \, 
    (q - s \, \nr_{1i} - t \, \nr_{1b}) \nonumber\\ 
    = c \times \textnormal{(product of even letters)} \times \textnormal{(further polynomials)} \, , 
\end{align}
where the further polynomials commonly are related to the function $t$. 
We have verified that denominators of several of these letters vanish if the further polynomials vanish. 
However, the resulting apparent poles are cancelled against zeroes of the numerators; these seem to be spurious singularities introduced by insisting to write the one-forms as $\mathrm{d}\log$'s. 
While this casts doubt on the paradigm that a $\mathrm{d}\log$-form is in all cases advantageous, we leave more detailed studies of this phenomenon to further work.

Despite this observation, to show explicitly the canonical form of the differential equation and for a first application, we attempt to rewrite these letters as $\mathrm{d}\log$'s. This requires explicit integration techniques. 
The basic strategy is the following: 
Integrating an entry of the matrix $A$ in eq.~\eqref{eq:MI:Deq} using a computer algebra system like \texttt{Mathematica}~14~\cite{Mathematica:2025}, we may observe that the primitive can be written in terms of a logarithm, confirming the $\mathrm{d}\log$-form of the entry. 
However, due to the complexity of the involved expressions, this direct approach does often not succeed.

An important step to simplify the problem is to note that using the algebraic identity eq.~\eqref{eq:NR:Rewrite} in the basis~\eqref{eq:MI:NR} and the resulting decomposition~\eqref{eq:NR:LetterDecomposition}, it is sufficient to determine differential forms $\alpha^{(a)}$ and $\alpha^{(b)}$ proportional to $\nr^{(a)}$ and $\nr^{(b)}$, respectively. 
The latter are more convenient for direct integration, since the interior square roots $r_4$ in these depend only on the variables $m^2$ and $s_1$. 
Therefore, direct integration of terms involving $\nr^{(a)}$ or $\nr^{(b)}$ with respect to all variables except for $m^2$ and $s_1$ becomes easier than an integration of terms involving $\nr_{1a}$ or $\nr_{1b}$.

As an example, assume that we are given an entry of the $A$-matrix in eq.~\eqref{eq:MI:Deq} in its general form, i.e.\ a one-form
\begin{align}
\label{eq:OneForm}
    \alpha = \sum_{i = 1}^7 \alpha_i \, \mathrm{d}x^i \, , 
\end{align}
where $\alpha_i$ are algebraic functions and we recall eq.~\eqref{eq:Kinematics:Vector} for the notation. 
This form can be directly derived from considering partial derivative matrices. 
Now, we may perform the split~\eqref{eq:NR:LetterDecomposition} at the level of partial derivatives, and find a decomposition of the form
\begin{align}
\label{eq:NR:Letters:ds12}
    \alpha = \alpha^{(a)} + \alpha^{(b)} = \frac{\sqrt{s_1} \, \nr^{(a)}(q + s \, r_4)}{t} \, \mathrm{d}s_{12} + \dots \, ,
\end{align}
where the ellipsis denotes terms either not proportional to $\mathrm{d}s_{12}$ or not proportional to $\nr^{(a)}$. 
n the case that $\alpha^{(a)}$ is independent of further square roots, the functions $q$, $s$ and $t$ can be chosen as polynomials. 
Then, we may integrate $\alpha^{(a)}$ with respect to $s_{12}$ and recognise that the primitive can be written in terms of a logarithm. 
Therefore, $\alpha^{(a)}$ can be written as a $\mathrm{d}\log$ form. Note that a linear combination of $\mathrm{d}\log$-forms is again a $\mathrm{d}\log$-form, and $\alpha^{(a)}$ is a $\mathrm{d}\log$-form if and only if $\alpha^{(b)}$ is. Consequently, also $\alpha$ can then be expressed in $\mathrm{d}\log$ form.

For letters involving further square roots, it is less clear whether they can be written in $\mathrm{d}\log$-form, since direct integration does still not exhibit logarithms at first sight. 
In principle, in such cases, instead we could try to integrate eq.~\eqref{eq:NR:Letters:ds12} by differentiating a suitable ansatz involving a logarithm. 
However, the resulting equations turn out to be too complicated to be solved easily. 
Nevertheless, we find $\mathrm{d}\log$-forms for all of these letters and describe our method in the following. 
As a somewhat easy example that can also be derived using different means, in Appendix~\ref{app:ReconstructLetters}, we also employ the method to determine the structure of the right-hand side of eq.~\eqref{eq:NR:LetterDecomposition:Example}.

To describe the derivation of these letters, assume that we are given a one-form decomposed as in eq.~\eqref{eq:NR:Letters:ds12}, where now the functions $q$ and $s$ also depend on further square roots. 
To proceed, we first need to verify whether $\alpha$ or equivalently $\alpha^{(a)}$ can be written in $\mathrm{d}\log$-form at all. 
For this, we consider the following quantity:
\begin{align}
\label{eq:NR:Letters:Integral}
    \widetilde{W} \equiv \exp\Bigg(\int_\gamma \alpha^{(a)}\Bigg) \, , 
\end{align}
where $\gamma$ is a path between rational-valued phase-space points. 
Now, assume that $\alpha^{(a)}=\mathrm{d}\log(W)$ for an algebraic function $W$, say of degree $\tilde{d}$. 
Then, up to branches,
\begin{align}
\label{eq:NR:Letters:Integral:Exp}
    \widetilde{W} = \frac{W(\gamma(1))}{W(\gamma(0))} \, .
\end{align}
which is an algebraic number of degree bounded by $\tilde{d}^2$. 
Put differently, if $\widetilde{W}$ admits a minimal polynomial, we conclude that $\alpha^{(a)}$ is a $\mathrm{d}\log$-form. 
Now, using a computer algebra system, it is often not hard to determine high-precision numerical values for $\widetilde{W}$, although such systems do not necessarily recognise logarithmic primitives. 
Therefore, we may apply an integer relation algorithm (e.g.\ \texttt{PSLQ}~\cite{Ferguson:1999}) to the family $\{1,\widetilde{W},\dots,\widetilde{W}^{n}\}$ for large enough $n$ to find a minimal polynomial.

Unfortunately, we do not know the degree $\tilde{d}$ a priori, and hence also not the required $n$. 
However, we may directly read off the degree $\tilde{d}_0$ of the partial derivatives in $\alpha^{(a)}$, see eq.~\eqref{eq:NR:Letters:ds12}. 
In this case, it is reasonable to assume that
\begin{align}
\label{eq:NR:Letters:Power}
    \alpha^{(a)} = c \times \mathrm{d}\log(w) = \mathrm{d}\log(w^c) \, ,
\end{align}
where $w$ is an algebraic function of degree $\tilde{d}_0$ and $c\in\mathbb{Q}$. 
In other words, the $\mathrm{d}\log$-argument is a rational power of an algebraic function of degree $\tilde{d}_0$. 
Therefore, it is sufficient to apply the \texttt{PSLQ} algorithm to families $\{1,\widetilde{W}^{1/c},\dots,\widetilde{W}^{\tilde{d}_0^2/c}\}$, where we guess $c$ to be a ``simple'' rational number and the number of elements in the family is merely as large as $\tilde{d}_0^2+1$. 
In our case, for each considered letter we found a relation for $c$ one out of the set $\{1, 1/2, 1/4\}$, and we prefer to write the resulting $\mathrm{d}\log$ as $c\times\mathrm{d}\log(w)$ instead of $\mathrm{d}\log(w^c)$.\footnote{However, since there is no rigorous degree bound, a possible failure of the algorithm does not necessarily imply that a $\mathrm{d}\log$-form for a certain letter does not exist. 
Furthermore, an integer relation-based strategy may always fail simply due to insufficient working precision. 
We note that applying the algorithm to the form $\omega^E$ of ref.~\cite{FebresCordero:2023pww}, we did not immediately find a minimal polynomial. 
This gives an argument that this form is indeed going beyond $\mathrm{d}\log$'s, but no proof.}

To further reduce $\tilde{d}_0$ and ease the application of the integer relation algorithm, we rationalise the root $r_4$, such that the partial derivatives only consider simple square roots. 
Note that the resulting minimal polynomial may not be expected to have small coefficients, making such simplifications important to keep the algorithm feasible.

Having verified that the considered one-form is a $\mathrm{d}\log$-form, as a next step, we would like to obtain the symbolic form of the argument $W$ (or $w$ in eq.~\eqref{eq:NR:Letters:Power}). 
To this end, we set all variables except for $s_{12}$ to fixed values $m_0^2,s_{1,0},\dots,s_{15,0}$, then repeatedly integrate eq.~\eqref{eq:NR:Letters:ds12} along straight-line paths parallel to the $s_{12}$-axis and exponentiate the integrals as in eq.~\eqref{eq:NR:Letters:Integral}. 
For example, we can choose these paths starting at $s_{12} = 0$ and ending at $s_{12} = b_k$ for rational numbers $b_k$, yielding as before values
\begin{align}
    \widetilde{W}_{0,k} \equiv \frac{W(m_0^2,s_{1,0},b_k,\dots)}{W(m_0^2,s_{1,0},0,\dots)} \, .
\end{align}
Using the \texttt{PSLQ} algorithm as described before, we may determine minimal polynomials of $\widetilde{W}_{0,k}$ over $\mathbb{Q}$. 
Then, by interpolation of rational functions, we can obtain the minimal polynomial of the univariate function
\begin{align}
    \widetilde{W}_0(s_{12}) \equiv \frac{W(m_0^2,s_{1,0},s_{12},\dots)}{W(m_0^2,s_{1,0},0,\dots)}
\end{align}
over $\mathbb{Q}(s_{12})$.
Given the minimal polynomial, we solve for its roots to get the function $\widetilde{W}_0$, and since the denominator is just a constant, on the univariate slice determined by $m_0^2,s_{1,0},\dots,s_{15,0}$ we can then set $W=\widetilde{W}_0$. 
To reconstruct the functional dependency of $W$ also on variables other than $s_{12}$, we either proceed analogously for multiple values $m_i^2,s_{1,i},\dots,s_{15,i}$ and again use interpolation, or we use the already reconstructed form of $\mathrm{d}\log(\widetilde{W}_0)$ to build an ansatz and fit coefficients in this ansatz by comparing it to eq.~\eqref{eq:NR:Letters:ds12}. 
At this point, given the restrictions obtained from the form of $\mathrm{d}\log(\widetilde{W}_0)$, the equations occurring in the ansatzing strategy become easy enough to be solved.

Using this strategy, we successfully determined letters involving both $\nr_{1a}$ and further square roots.
This enables us to write all of our differential equations in terms of $\mathrm{d}\log$-forms.

Finally, we note a technical detail about the reconstruction procedure of the involved rational functions. 
One could obtain the components $\alpha_i$ in eq.~\eqref{eq:OneForm} by symbolic manipulations or standard interpolation of numerical samples for the differential equations, however this requires a larger number of samples than necessary. 
In this work, we determine first the rational differential equations for the associated almost-canonical basis and then perform the basis change introducing square roots symbolically. 
To obtain the rational equations, we first guess their denominators from $\mathbb{Q}$-valued samples (see e.g.\ ref.~\cite{Heller:2021qkz}), such that merely numerators remain to be reconstructed. 
These may be found using only interpolation of the first five variables, while setting the last two to numerical values. 
The symbolic dependence on the last variables can subsequently be guessed by employing an integer relation algorithm.

\section{Differential equations and permutation closure}
\label{sec:DeqsPermutations}
Once we have found the alphabet, it remains to determine the $\mathbb{Q}$-valued matrices $A_k$ in eq.~\eqref{eq:MI:Deq} to get the full functional form of the differential equations. 
This can be achieved by sampling the differential equations a relatively small number of times and solving a linear system~\cite{Abreu:2018rcw}.

In this work, we first consider the differential equations for the identity permutation of the penta-box and hexa-box, which contain $148$ and additional $50$ letters, respectively. 
We find that nine of the letters do not appear in the $\varepsilon$-expansion of the master integrals up to weight $4$.

Afterwards, we permute the $\varepsilon$-factorising bases and their differential equations by all permutations of the external massless legs, which requires the introduction of further letters and square roots. 
In order to obtain a linearly independent alphabet sufficient to describe all permutations, we first compute the permutation closure of the alphabet in the identity permutation and detect linear dependencies between letters using the integer relation algorithm of \texttt{Pari/GP}~\cite{Pari:2025}. 
Then, we remove redundant ones to obtain a basis.

When deriving permuted differential equations with nested square roots, we also note a small caveat. 
Naively, one would assume that the branches of nested square roots are entirely free to choose and drop out of predictions for scalar integrals, as long as they are correctly analytically continued along the integration path. 
However, care must be taken concerning the relation~\eqref{eq:NR:Relation}. 
We compute the differential equations first at a base-point $\vec{x}_0$, where eq.~\eqref{eq:NR:Relation} holds for all square roots on their principal branches. 
However, at other points, for all square roots on their principal branches, eq.~\eqref{eq:NR:Relation} may only hold up to a sign. 
Thus, at these points we need to enforce that one of the square roots is not evaluated on its principal branch, but on the negative of its principal branch. 
This is automatically ensured by analytic continuation along an integration path starting at $\vec{x}_0$, but need not be satisfied if we choose different base-points for numerical integration, or if we permute the differential equations to obtain equations for crossed integral families.

In practice, we fix the resulting signs in an ad-hoc-manner by requiring that results for scalar master integrals are correctly obtained. 
This needs to be done only once for each permutation.

\section{Special function basis and numerical evaluation}
\label{sec:FunctionBasis}
An $\varepsilon$-factorised differential equation as eq.~\eqref{eq:MI:Deq} can be directly solved order-by-order in $\varepsilon$ in terms of iterated integrals. 
In this work, to remove redundancies for different permutations, we further improve the representation of $\varepsilon$-coefficients of master integrals by promoting an algebraically independent set of linear combinations thereof to transcendental basis functions. 
Remaining ones are then expressed as graded polynomials in these functions and zeta values~\cite{Gehrmann:2018yef, Chicherin:2020oor, Chicherin:2021dyp, Abreu:2023rco, Badger:2025ljy}. 
For example, for the weight-$3$- or $\varepsilon^3$-coefficient $I_{i,\sigma}^{(3)}$ of a master integral $I_i$ in a permutation $\sigma$, we write
\begin{align}
\label{eq:MI:FromBasisfunctions}
    I_{i,\sigma}^{(3)} = \sum_{j,k,l} a_{j,k,l} \, f_{1,j} \, f_{1,k} \, f_{1,l} + \sum_{j,k} b_{j,k} \, f_{1,j} \, f_{2,k} + \sum_j c_j \, f_{3,j} + \sum_j d_j \, f_{1,j} \, \zeta_2 + e \, \zeta_3 \, , 
\end{align}
where $a_{j,k,l}$, $b_{j,k}$, $c_j$, $d_j$ and $e$ are rational numbers depending on $i$ and $\sigma$, $\zeta_k \equiv \zeta(k)$ are values of the Riemann zeta function and $f_{w,j}$ denotes the $j$-th transcendental basis function of weight $w$. 
Thus, we have
\begin{align}
\label{eq:Functions:Definition}
    f_{w,j} = \sum_{k,\sigma} a_{k,\sigma} \, I_{k,\sigma}^{(w)}
\end{align}
for rational numbers $a_{k,\sigma}$ (depending on $w$ and $j$). 
In practice, for weights greater than one, only one $a_{k,\sigma}$ equals one and all others are zero. 
We build on the transcendental function basis defined in ref.~\cite{Asteriadis:2026olo} for the irreducible one-loop VBF pentagon topology. 
We give the numbers of basis functions at each weight in Table~\ref{tab:Statistics}, together with the numbers of square roots and letters when considering all permutations simultaneously.

\begin{table}[t]
\centering
\begin{tabular}{lccc}
\toprule
 & Pentagon~\cite{Asteriadis:2026olo} & $ + \, $PB & $ + \, $HB \\
\midrule
Square roots & $11$ & $ + \, 55 \, (\textnormal{$12$ nested})$ & $ + \, 6$ \\
Letters & $195$ & $ + \, 534$ & $ + \, 72$ \\
Weight-$1$ functions & $18$ & $ + \, 0$ & $ + \, 0$ \\
Weight-$2$ functions & $60$ & $ + \, 7$ & $ + \, 18$ \\
Weight-$3$ functions & $84$ & $ + \, 273$ & $ + \, 108$ \\
Weight-$4$ functions & $84$ & $ + \, 738$ & $ + \, 186$ \\
\bottomrule
\end{tabular}
\caption{Numbers of square roots, letters, and basis functions arising in the differential equations for various topologies, considering all permutations simultaneously.}
\label{tab:Statistics}
\end{table}

The decomposition~\eqref{eq:MI:FromBasisfunctions} is obtained order-by-order in the weight (see e.g.\ refs.~\cite{Gehrmann:2018yef, Chicherin:2020oor, Chicherin:2021dyp, Abreu:2023rco}). 
An explicit example at weight $3$ is given in Appendix~\ref{app:FunctionDecomposition}. 
At each weight $w$, we first determine a set of linearly independent coproducts of weight-$w$-coefficients of master integrals~\cite{Goncharov:2005sla, Goncharov:2010jf, Duhr:2011zq, Duhr:2012fh, vonManteuffel:2013vja} by row reduction; to this end, we use \texttt{Reduze~2}. 
Here, we drop all coproducts that can be obtained by multiplying functions of lower weight and zeta values, which are at this point already known. 
The $\varepsilon$-coefficients corresponding to the independent coproducts are then defined to be new basis functions at weight $w$. 
Afterwards, we express the coproducts of all master integrals in terms of the coproducts of these basis functions (and coproducts of products of lower-weight functions) and obtain the decomposition~\eqref{eq:MI:FromBasisfunctions} up to primitive elements of weight $w$. 
Namely, this may still miss the last term in eq.~\eqref{eq:MI:FromBasisfunctions}, i.e.\ rational multiples of transcendental constants of weight $w$. 
Note that it is a consequence of our choice of basis functions that all explicit transcendental constants arising are in fact zeta values~\cite{Abreu:2023rco}. 
Such terms may be determined by computing values of master integrals at a single phase-space point and applying a \texttt{PSLQ} algorithm.

From the $\varepsilon$-factorised differential equation~\eqref{eq:MI:Deq} and the definition~\eqref{eq:Functions:Definition}, it is easy to obtain an $\varepsilon$-independent differential equation for the vector $\vec{f} \equiv (f_{0,1},f_{1,1},\dots,f_{4,1008})^T$ of basis functions. 
Here, we include a function $f_{0,1} \equiv 1$ for convenience. 
The equation can be written as
\begin{align}
\label{eq:Functions:Deq}
    \mathrm{d}\vec{f} = \sum_k\alpha_k \, B_k \, \vec{f}_\mathrm{aug} \, , 
\end{align}
where $\alpha_k$ are the letters derived in Section~\ref{sec:Alphabet} and $\vec{f}_\mathrm{aug}$ is the vector $\vec{f}$ augmented with e.g.\ weight-$3$ products of three weight-$1$ functions. 
The matrices $B_k$ are $\mathbb{Q}$-valued and only couple functions of weight $w$ onto functions of weight $w - 1$. 
Therefore, upon choosing a base-point $\vec{x}^\prime$ and determining $\vec{f}(\vec{x}^\prime)$ by other means, we obtain values at an arbitrary point $\vec{x}$ via
\begin{align}
\label{eq:Functions:Deq:Integrated}
    \vec{f}(\vec{x}) = \vec{f}(\vec{x}^\prime) + \int_\gamma \, \, \sum_k \alpha_k \, B_k \, \vec{f}_\mathrm{aug} \, , 
\end{align}
where $\gamma$ is a path from $\vec{x}^\prime$ to $\vec{x}$. 
We compute the integral numerically (see e.g.\ refs.~\cite{Caffo:1998du, Boughezal:2007ny, Czakon:2020vql, Badger:2025ljy, Czakon:2026tog}), using the \texttt{Odeint} library from \texttt{Boost~C++}~\cite{Boost:2026}.

Thus, to specify the solution, in addition to the differential equations~\eqref{eq:Functions:Deq} we are required to provide boundary values $\vec{f}(\vec{x}^\prime)$. 
We pick
\begin{align}
\label{eq:xprime}
    \frac{\vec{x}^\prime}{\mu_0^2} \equiv 
    \bigg(1 - \frac{27 \, i}{256},
    \frac{11}{4} + \frac{251 \, i}{512},
    \frac{13}{2} + \frac{149 \, i}{512},
    \frac{7}{4} + \frac{7 \, i}{64},
    - 2 + \frac{19 \, i}{512},
    \frac{25}{2} + \frac{73 \, i}{512},
    - \frac{15}{4} + \frac{15 \, i}{64}\bigg)^T\,,
\end{align}
such that $\Re(\vec{x}^\prime)$ lies in the physical $s_{45}$-scattering region~\eqref{eq:s45Region}. 
However, we choose $\vec{x}^\prime$ with non-zero imaginary parts of kinematic invariants compatible with Feynman's $i\delta$ prescription. 
Then, if we want to obtain values for $\vec{f}$ at a phase-space point $\vec{x}$ with real-valued invariants, we compute the integral~\eqref{eq:Functions:Deq:Integrated} along a straight-line path from $\vec{x}^\prime$ to $\vec{x}$. 
With our choice of base-point, we ensure that the imaginary parts of kinematic invariants vanish only at the very end of the path, thereby circumventing singularities.

To obtain boundary values $\vec{f}(\vec{x}^\prime)$, we first compute values for scalar master integrals at $\Re(\vec{x}^\prime)$ using \texttt{AMFlow}~\cite{Liu:2022chg} and then change the basis to the canonical one, obtaining values $\vec{I}(\Re(\vec{x}^\prime))$. 
At this point, we note that even if we choose $\Re(\vec{x}^\prime)$ such that all canonical integrals are regular, this need not necessarily hold for the basis change. 
Therefore, we use l'Hôpital's rule to evaluate canonical master integrals whose basis change from scalar masters is singular. 
Afterwards, we transport the values at $\Re(\vec{x}^\prime)$ using the differential equations~\eqref{eq:MI:Deq} to $\vec{x}^\prime$, employing series expansion techniques~\cite{Moriello:2019yhu}. 
Finally, at $\vec{x}^\prime$, we use the definition~\eqref{eq:Functions:Definition} to translate $\vec{I}(\vec{x}^\prime)$ into values $\vec{f}(\vec{x}^\prime)$.\footnote{Although \texttt{AMFlow} would in principle also be applicable directly at the complex-valued point $\vec{x}^\prime$, this would require symbolic IBP reduction tables which we found expensive to generate. 
We note that the computational load would remain within the capabilities of our setup if we parametrised the problem such that there is only a single complex variable, but the above method works as well.} 
As a consistency check, we derive the constants up to weight $2$ also by requiring regularity at a degenerate phase-space point in the Euclidean region.

Although we avoid poles on the integration contour by our choice of base-point, we still need to make sure that spurious square-root discontinuities are avoided. 
To this end, we use the automated analytic continuation algorithm of ref.~\cite{Asteriadis:2026olo}. 
This is based on the fact that potential branch cut-crossings can be detected by solving univariate polynomial equations in the path variable, which we do using the \texttt{FLINT} library~\cite{flint:2025}. 
This algorithm has originally been developed for the case of simple square roots, but generalises easily to the nested square roots required for this work. 
For details, see Appendix~\ref{app:AnalyticContinuation}.

In order to speed up the numerical evaluations, it is useful to optimise the computation of letters as far as possible. 
In particular, following ref.~\cite{Badger:2025ljy}, we scan for the most convenient variables for each polynomial occurring in them. 
Moreover, we compute all polynomials simultaneously, detecting common subexpressions and reducing the number of floating-point operations with the \texttt{FORM} optimiser~\cite{Kuipers:2013pba}. 
We implement the master integrals in double and quadruple precision using \texttt{GCC~libquadmath}~\cite{Quadmath:2024}. 
As an example, in Table~\ref{tab:Benchmark}, we give benchmark values for the almost-canonical top-sector master integrals~\eqref{eq:MI:TopSector}, at the randomly chosen phase-space point
\begin{align}
\label{eq:Benchmark:Point}
    \frac{\vec{x}_1}{\mu_0^2} = \bigg(\frac{87}{34},\frac{7}{38},\frac{17}{37},\frac{415301}{549746}, - \frac{1105}{1702},\frac{44}{23}, - \frac{88093}{289731}\bigg)^T \, .
\end{align}
In ancillary files, we give results for all almost-canonical master integrals in all permutations at the same point.

We note that since our code computes all master integrals in all permutations, we also obtain values for these integrals at permuted phase-space points $\sigma(\vec{x}_1)$. 
In this way, we may estimate the errors by computing the same master integrals either in the identity permutation, or in a permuted variant at an accordingly permuted phase-space point. 
Using quadruple precision, we may reach a relative error estimate of less than $10^{-16}$.

\begin{table}[t]
\centering
\scriptsize
\begin{tabular}{cccccc}
\toprule
& $\varepsilon^0$ & $\varepsilon^1$ & $\varepsilon^2$ & $\varepsilon^3$ & $\varepsilon^4$ \\
\midrule
$\frac{\mu_0^4}{r_2} \, \IPB_{109}$ & $0$ & $0$ & $0.2890499621527367351$ & $1.544728505517956330$ & $6.813371739655722401$ \\[4pt]
$\frac{r_1}{\mu_0^4} \, \IPB_{110}$ & $0$ & $0$ & $0$ & $0$ & $0$ \\[4pt]
$\frac{r_1}{\mu_0^4} \, \IPB_{111}$ & $0$ & $0$ & $0$ & $0.01338274386848353513$ & $0.08004010605728960091$ \\[4pt]
$\frac{\mu_0^4}{r_2} \, \IHB_{64}$ & $0$ & $0$ & $0$ & $1.134957825733761948$ & $8.927196394039801702$ \\[4pt]
$\frac{r_1}{\mu_0^4} \, \IHB_{65}$ & $0$ & $0$ & $0$ & $0$ & $ - 0.0007359003524335330293$ \\[4pt]
$\frac{r_1}{\mu_0^4} \, \IHB_{66}$ & $0$ & $0$ & $0$ & $0$ & $ - 0.00008371400734877308598$ \\
\bottomrule
\end{tabular}
\caption{Values for almost-canonical top-sector master integrals eq.~\eqref{eq:MI:TopSector} at the phase-space point eq.~\eqref{eq:Benchmark:Point}. 
All numbers smaller in magnitude than $10^{-20}$ are rounded to zero.}
\label{tab:Benchmark}
\end{table}

\section{Conclusions}
\label{sec:Conclusions}
In this work, we presented a calculation of two-loop penta-box and hexa-box Feynman integral families required for non-factorisable QCD corrections to Higgs-boson production via vector boson fusion.
These planar and non-planar integral families have five legs, of which one is massive, and eight propagators, of which two are massive.
We derived basis integrals satisfying $\varepsilon$-factorised differential equations, which was complicated by seven-variable kinematics and nested square roots.
The results for the integral basis are provided in computer-readable form in the ancillary files accompanying this article on \texttt{arXiv}.
This is the first time a canonical basis was constructed for non-planar Feynman integrals depending on seven scales.

To facilitate future applications, we expressed the integrals for all permutations of the massless legs in terms of a common basis of algebraically independent functions.
For example, the same transcendental function basis may also be used to compute two-loop QCD corrections to the Higgs-Strahlung process.
We expect these results to allow for compact representations of amplitudes and analytical cancellations of ultraviolet and infrared poles in finite remainders.
We evaluate the integrals by solving the differential equations numerically, benefitting from their simpler form due to our analytical preparations.

We have put particular focus on the nested square roots involved in the penta-box topology and described strategies for deriving master integrals, letters, and analytic continuations for such cases. 
We believe that these will be more generally applicable for similar problems, which we expect to become increasingly common for integrals with massive propagators.

In this work, we calculated only two of the two-loop families required for VBF at NNLO, which are shown in the first row of Figure~\ref{fig:VBF2Loop}.
The complete two-loop amplitudes involve two additional double pentagon families shown in the second row of Figure~\ref{fig:VBF2Loop}, resembling the topologies DPmz and DPzz in the notation of ref.~\cite{Abreu:2023rco}.
We expect these to be more difficult, in particular encompassing further nested square roots and elliptic structures.
For this reason, we leave them for future work.

\acknowledgments
We are grateful to Konstantin Asteriadis for many helpful discussions.
The Feynman diagrams in this work were drawn with \texttt{TikZ-Feynman}~\cite{Ellis:2016jkw}.

\appendix
\section{Example for reconstruction of \texorpdfstring{\boldmath $\mathrm{d}\log$}{dlog}-letters with nested square roots}
\label{app:ReconstructLetters}
As an illustration for the reconstruction of letters described in Subsection~\ref{subsec:Alphabet:WithNested:Derivation}, let us re-derive the argument of the first term of eq.~\eqref{eq:NR:LetterDecomposition:Example}, given only partial derivative information as in eq.~\eqref{eq:NR:Letters:ds12}. 
We denote the one-form by $\alpha^{(a)}$ and rationalise the root $r_4$ using a Landau variable
\begin{align}
    s_1 = m^2\frac{(1 + x)^2}{x} \, .
\end{align}
We obtain
\begin{align}
\label{eq:NR:Letters:ds12:Example}
    \alpha^{(a)} = \frac{2 \, m^2 s_{34} (1 + x)^2 + s_{15} x ( - s_{12} + s_{15} x - s_{34} (1 + x))}{(m^2 s_{34} - s_{12} s_{15} x + 2 \, m^2 s_{34} x + m^2 s_{34} x^2) \, \sqrt{x/m^2} \, \nr^{(a)}} \, \mathrm{d}s_{12} + \dots \, , 
\end{align}
where the ellipsis denotes terms independent of $\mathrm{d}s_{12}$ and
\begin{align}
    \nr^{(a)} & = \sqrt{\frac{m^2}{x}} \times \big(s_{12}^2 + 4 \, m^2 s_{34} - 2 \, s_{12} s_{34} + s_{34}^2 - 2 \, s_{12} s_{15} x + 8 \, m^2 s_{34} x - 2 \, s_{12} s_{34} x \nonumber\\
    & \hspace{11pt} - 2 \, s_{15} s_{34} x + 2 \, s_{34}^2 x + s_{15}^2 x^2 + 4 \, m^2 s_{34} x^2 - 2 \, s_{15} s_{34} x^2 + s_{34}^2 x^2\big)^{1/2}
\end{align}
expressed in terms of the Landau variable. 
We set all variables except for $s_{12}$ to fixed values
\begin{align}
\label{eq:NR:Letters:ds12:Example:FixedValues}
    x_0 = 1 \, , \quad m_0^2 = 2 \, , \quad s_{23,0} = 3 \, , \quad s_{34,0} = 4 \, , \quad s_{45,0} = 5 \, , \quad  s_{15,0} = 6 \, .
\end{align}
Then, we may easily integrate eq.~\eqref{eq:NR:Letters:ds12:Example} numerically with respect to $s_{12}$ from $0$ to $1$. 
Exponentiating the integral as in eq.~\eqref{eq:NR:Letters:Integral}, we get a value $\widetilde{W}_{0,1}$.
Since the partial derivatives only contain a simple square root, we expect that a rational power of $\widetilde{W}_{0,1}$ admits a minimal polynomial of degree $4$. 
Indeed, using a \texttt{PSLQ} algorithm, we find that to $1000$ digits precision,
\begin{align}
    \widetilde{W}_{0,1}^4 - \frac{901}{208} \, \widetilde{W}_{0,1}^3 + \frac{72153}{10816} \, \widetilde{W}_{0,1}^2 - \frac{901}{208} \, \widetilde{W}_{0,1} + 1 = 0 \, .
\end{align}
We repeat this exercise, integrating with respect to $s_{12}$ from $0$ to values
\begin{align}
    \bigg\{\frac{1}{2},\frac{3}{4},1,\frac{5}{4},\frac{3}{2},\frac{7}{4},\frac{9}{4},\frac{5}{2},\frac{11}{4},3,\frac{7}{2},\frac{9}{2},5\bigg\} \, .
\end{align}
These values have been chosen such that the integrals are real and the \texttt{PSLQ} algorithm recognises the desired polynomial relation that we want to reconstruct. 
Using Thiele's interpolation formula (\cite{Thiele:1909}, Section 40), we then find that
\begin{align}
    & \widetilde{W}_0^4 + b \, \widetilde{W}_0^3 + c \, \widetilde{W}_0^2 + b \, \widetilde{W}_0 + 1 = 0 \, ,  \nonumber\\
    \textnormal{where} \quad \widetilde{W}_0 & = \exp\Bigg(\int_0^{s_{12}}\alpha_0^{(a)\prime} \, \, \mathrm{d}s_{12}^\prime\Bigg) \, , \nonumber\\
    b & = \frac{17 \, (s_{12}^2 - 16 \, s_{12} + 68)}{16 \, (3 \, s_{12} - 16)} \, , \nonumber\\
    c & = \frac{16 \, s_{12}^4 - 512 \, s_{12}^3 + 7721 \, s_{12}^2 - 50272 \, s_{12} + 115200}{64 \, (3 \, s_{12} - 16)^2} \, .
\end{align}
Here, $\alpha_0^{(a)\prime}$ denotes the $\mathrm{d}s_{12}$-coefficient of the one-form $\alpha^{(a)}$ in eq.~\eqref{eq:NR:Letters:ds12:Example}, with all variables except for $s_{12}$ set to values~\eqref{eq:NR:Letters:ds12:Example:FixedValues} and $s_{12}$ replaced by an integration variable $s_{12}^\prime$. 
This equation is solved by \texttt{Mathematica} in terms of nested radicals, which however can again be denested. 
For example, one solution to it is given by
\begin{align}
    \widetilde{W}_0 & = - \frac{17 + \sqrt{33}}{64 \, (3 \, s_{12} - 16)} \, \bigg(s_{12}^2 - 16 \, s_{12} + 68 + (s_{12} - 2) \, \sqrt{s_{12}^2 - 28 \, s_{12} + 132}\bigg).
\end{align}
One easily verifies that e.g.\ near $s_{12} = 1$ and all square roots taken on their principal branch, we have
\begin{align}
    \widetilde{W}_0 = \frac{W_0(s_{12})}{W_0(s_{12} = 0)} \, 
\end{align}
where $W_0$ is the $\mathrm{d}\log$-argument of the first term on the right-hand side of eq.~\eqref{eq:NR:LetterDecomposition:Example}, with values~\eqref{eq:NR:Letters:ds12:Example:FixedValues} substituted for all variables except $s_{12}$.

\section{Example for decomposition into basis functions}
\label{app:FunctionDecomposition}
As an example for the decomposition procedure described in Section~\ref{sec:FunctionBasis}, let us consider the integrals $\IPB_3$ and $\IPB_4$ in the identity permutation. 
Their weight-$2$-solutions are given by
\begin{align}
    I^{\mathrm{PB},(2)}_3 & = - f_{1,1}f_{1,6} + f_{1,6}^2 - f_{2,11} \, , \nonumber\\
    I^{\mathrm{PB},(2)}_4 & = - 2 \, f_{1,1}^2 - f_{2,11} - 3 \, \zeta_2 \, , 
\end{align}
and their differential equations read
\begin{align}
    \mathrm{d}\IPB_3 & = \varepsilon \, (( - \mathrm{d}\log(W_1) + \mathrm{d}\log(W_4) - 2 \, \mathrm{d}\log(W_9)) \, \IPB_3 \nonumber\\
    & \hspace{11pt} + ( - \mathrm{d}\log(W_1) + \mathrm{d}\log(W_9)) \, \IPB_4) \, , \nonumber\\
    \mathrm{d}\IPB_4 & = \varepsilon \, (( - 4 \, \mathrm{d}\log(W_4) + 4 \, \mathrm{d}\log(W_9)) \, \IPB_{3} 
    - 2 \, \mathrm{d}\log(W_9) \, \IPB_4) \, ,
\end{align}
for letters $\mathrm{d}\log(W_i)$ whose explicit form is not relevant in the following.
Therefore, for the $(1,1,1)$ iteration of the coproducts at weight $3$, we find
\begin{align}
    \Delta_{1,1,1}(I^{\mathrm{PB},(3)}_3) & = 2 \, \Delta_{1,1}(f_{1,1}^2)\otimes\log(W_{1}) - 2 \, \Delta_{1,1}(f_{1,1}^2)\otimes\log(W_{9}) \nonumber\\
    & \hspace{11pt} + \Delta_{1,1}(f_{1,1} f_{1,6})\otimes\log(W_{1}) - \Delta_{1,1}(f_{1,1} f_{1,6})\otimes\log(W_{4}) \nonumber\\
    & \hspace{11pt} + 2 \, \Delta_{1,1}(f_{1,1} f_{1,6})\otimes\log(W_{9}) - \Delta_{1,1}(f_{1,6}^2)\otimes\log(W_{1}) \nonumber\\
    & \hspace{11pt} + \Delta_{1,1}(f_{1,6}^2)\otimes\log(W_{4}) - 2 \, \Delta_{1,1}(f_{1,6}^2)\otimes\log(W_{9}) \nonumber\\
    & \hspace{11pt} + 3 \, \Delta_{1,1}(f_{2,11})\otimes\log(W_{1}) + \Delta_{1,1}(f_{2,11})\otimes\log(W_{4}) \nonumber\\
    & \hspace{11pt} - 6 \, \Delta_{1,1}(f_{2,11})\otimes\log(W_{9}) \, , \nonumber\\
    \Delta_{1,1,1}(I^{\mathrm{PB},(3)}_4) & = 4 \, \Delta_{1,1}(f_{1,1}^2)\otimes\log(W_{9}) + 4 \, \Delta_{1,1}(f_{1,1} f_{1,6})\otimes\log(W_{4}) \nonumber\\
    & \hspace{11pt} - 4 \, \Delta_{1,1}(f_{1,1} f_{1,6})\otimes\log(W_{9}) - 4 \, \Delta_{1,1}(f_{1,6}^2)\otimes\log(W_{4}) \nonumber\\
    & \hspace{11pt} + 4 \, \Delta_{1,1}(f_{1,6}^2)\otimes\log(W_{9}) - 4 \, \Delta_{1,1}(f_{2,11})\otimes\log(W_{4}) \nonumber\\
    & \hspace{11pt} + 12 \, \Delta_{1,1}(f_{2,11})\otimes\log(W_{9}) \, , 
\end{align}
where $(1,1)$-coproducts of weight-$2$ functions may also be expressed in terms of logarithms of letters. 
In addition, we consider the part of the $(2,1)$-iteration of the coproducts which depends on $\zeta_2$, that is
\begin{align}
    \Delta_{2,1}(I^{\mathrm{PB},(3)}_3)|_{\zeta_2} & = 3 \, \zeta_2\otimes\log(W_1) - 3 \, \zeta_2\otimes\log(W_9) \, , \nonumber\\
    \Delta_{2,1}(I^{\mathrm{PB},(3)}_4)|_{\zeta_2} & = 6 \, \zeta_2\otimes\log(W_9) \, .
\end{align}
Using row reduction, we find that $\Delta_{1,1,1}(I^{\mathrm{PB},(3)}_3) + \Delta_{2,1}(I^{\mathrm{PB},(3)}_3)|_{\zeta_2}$ can not be reduced onto objects which are already known. 
Therefore, we declare a new basis function $f_{3,85} \equiv I^{\mathrm{PB},(3)}_3$. On the other hand, we find the relation
\begin{align}
    \Delta_{1,1,1}(I^{\mathrm{PB},(3)}_4) + \Delta_{2,1}(I^{\mathrm{PB},(3)}_4)|_{\zeta_2} & = \Delta_{1,1,1}\bigg(\frac{2}{3} \, f_{1,1}^3 + 2 \, f_{1,1}^2 f_{1,6} - \frac{4}{3} \, f_{1,6}^3 + 4 \, f_{1,6} f_{2,11} \nonumber\\
    & \hspace{11pt} - 4 \, f_{3,11} + 2 \, \zeta_2 \, f_{1,1} + 4 \, \zeta_2 f_{1,6}\bigg) + \Delta_{2,1}(\dots)|_{\zeta_2} \, , 
\label{eq:FunctionDecomposition:Example}
\end{align}
where the second bracket on the right-hand side contains the same terms as the first and $f_{3,11}$ is already known from one loop~\cite{Asteriadis:2026olo}.
Since we only considered $(1,1,1)$ and $(2,1)$-coproducts and not the function itself, the elements on the right-hand side of eq.~\eqref{eq:FunctionDecomposition:Example} give a functional decomposition of $I^{\mathrm{PB},(3)}_4$ only up to rational multiples of $\zeta_3$. 
We match these multiples by explicit evaluation and obtain in total
\begin{align}
    I^{\mathrm{PB},(3)}_3 & = f_{3,85} \, , \nonumber\\
    I^{\mathrm{PB},(3)}_4 & = \frac{2}{3} \, f_{1,1}^3 + 2 \, f_{1,1}^2 f_{1,6} - \frac{4}{3} \, f_{1,6}^3 + 4 \, f_{1,6} f_{2,11} - 4 \, f_{3,11} + 2 \, \zeta_2 \, f_{1,1} \nonumber\\
    & \hspace{11pt} + 4 \, \zeta_2 f_{1,6} + \frac{8}{3} \, \zeta_3 \, .
\end{align}

\section{Algorithmic analytic continuation of nested square roots}
\label{app:AnalyticContinuation}
In this appendix, we give details on the algorithm for the analytic continuation of (nested) square roots employed in the numerical integration strategy of Section~\ref{sec:FunctionBasis}, see also ref.~\cite{Asteriadis:2026olo}.

The algorithm is based on the observation that if we already know the analytic continuation of a function $W(t) \equiv W(\gamma(t))$ of a real variable $t$ along a path $\gamma$, and also a necessary condition for the vanishing of $\Im(W(t))$ that is satisfied only for finitely many points, we may analytically continue the square root $\sqrt{W(t)}$ (where $\sqrt{\cdot}$ denotes the principal branch).
Indeed, then it is sufficient to compute these points and verify at each of them numerically whether $W(t)$ is negative real and crosses the branch cut. 
If this is the case, we may multiply $\sqrt{W(t)}$ with appropriate signs to obtain an analytic continuation.

In the case that we are considering a simple square root and thus $W(t)$ is a polynomial in $t$ (possibly with complex-valued coefficients), $\Im(W(t))$ is a polynomial in $t$ and it is straightforward to determine such necessary conditions by searching for the real roots of $\Im(W(t))$. 
On the other hand, if $W(t) = \nr_{1a/b}^2(t)$ involves itself square roots, we are dealing with a nested square root. 
Nevertheless, we can find a polynomial $Q(t)$ such that if $W(t)$ crosses the branch cut, $Q(t)$ vanishes. 
In order to see this, consider first the minimal polynomial $P \in \mathbb{R}(t)[X]$ of $W(t)$, such that for all $t \in \mathbb{R}$, we have
\begin{align}
    P(t)(W(t)) = 0 \, .
\end{align}
In fact, it is not hard to see that the coefficients of $P(t)$ are themselves polynomials in $t$, as opposed to rational functions. 
Then, for all $t \in \mathbb{R}$, $P(t) \in \mathbb{R}[X]$ and thus the non-real roots of $P(t)$ come in pairs of complex conjugates. 
In particular, we see that also $\overline{W(t)}$ is a root of $P(t)$. 
Now, if $W(t)$ crosses the branch cut on the negative real axis for some $t_0 \in \mathbb{R}$, this implies, by the continuous dependency of polynomial roots on parameters, that $P(t_0)$ must have a double zero. To justify this, note that $W(t)$ ``crossing the branch cut at $t_0$'' implies in particular that it is non-real for $t\neq t_0$ in a neighbourhood of $t_0$. 
Hence, two distinct roots collide at $t = t_0$.

Thus, if $W(t)$ crosses the negative real axis at $t_0$, we necessarily have $\Delta(P(t_0)) = 0$, where $\Delta$ denotes the discriminant of a polynomial. 
Hence, $\Delta(P(t))$, which is a polynomial in $t$, has the desired property that it vanishes if $\Im(W(t))$ vanishes. 
Further factorising this polynomial, we can show that to find branch cut crossings of the nested square root, it is sufficient to examine points where the following polynomial vanishes:
\begin{align}
\label{eq:AnalyticContinuation:Polynomial}
    Q(t) \equiv 4 \, \Im(P_b(t))^4 - \Im(\nr_{1i}^2(t))^2 + 4 \, \Re(\nr_{1i}^2(t)) \, \Im(P_b(t))^2 \, .
\end{align}
Additionally, we test for zeroes of $\Im(\nr_{1i}^2(t))$ to detect branch cut crossings of the inner square root.

\bibliographystyle{JHEP}
\bibliography{literature.bib}

\end{document}